\documentclass[10pt]{article}
\usepackage[utf8]{inputenc}
\usepackage{lipsum}
\usepackage{geometry}
\usepackage{authblk}
\usepackage{verbatim}
\usepackage{amsmath}
\usepackage{amssymb}
\usepackage{amsfonts}
\usepackage{array, etoolbox}
\usepackage{longtable}
\usepackage{multirow}
\usepackage{multicol}
\usepackage{caption}
\usepackage{subcaption}
\usepackage{adjustbox}
\usepackage[dvipsnames]{xcolor}
\usepackage{hyperref}
\usepackage{url}
\usepackage{accents}
\usepackage{dsfont}
\usepackage{bbm}
\usepackage{subfiles}
\usepackage{pdflscape}
\newcommand*\underdot[1]{%
  \underaccent{\dot}{#1}}
\hypersetup{
    colorlinks=true,
    linkcolor=blue,
    filecolor=magenta,      
    urlcolor=cyan,
}
\newcolumntype{P}[1]{>{\centering\arraybackslash}p{#1}}
\preto\tabular{\setcounter{magicrownumbers}{-1}}
\newcounter{magicrownumbers}

\graphicspath{{images/}{../images/}}
\title{Causal Analysis of Carnatic Music
: A Preliminary Study}
\author[1,*]{Abhsihek Nandekar}
\author[1,2]{Preeth Khona}
\author[1]{Rajani M.B.}
\author[3,4]{Anindya Sinha}
\author[1]{Nithin Nagaraj}
\affil[1]{\footnotesize School of Humanities, National Institute of Advanced Studies, Bengaluru, India}
\affil[2]{\footnotesize Manipal Academy of Higher Education, Manipal, India}
\affil[3]{\footnotesize School of Natural Sciences and Engineering, National Institute of Advanced Studies, Bengaluru, India}
\affil[4]{\footnotesize College of Humanities,
University of Exeter,
Exeter, United Kingdom}
\affil[*]{abhisheknandekar4@gmail.com}

\date{}

\begin{document}

\maketitle

\begin{abstract}
The musicological analysis of Carnatic music is challenging, owing to its rich structure and complexity. Automated \textit{r\=aga} classification, pitch detection, tonal analysis, modelling and information retrieval of this form of southern Indian classical music have, however, made significant progress in recent times. A causal analysis to investigate the musicological structure of Carnatic compositions and the identification of the relationships embedded in them have never been previously attempted. In this study, we propose a novel framework for causal discovery, using a compression-complexity measure. Owing to the limited number of compositions available, however, we generated surrogates to further facilitate the analysis of the prevailing causal relationships. Our analysis indicates that the context-free grammar, inferred from more complex compositions, such as the \textit{M\=e\d{l}akarta} \textit{r\=aga}, are a \textit{structural cause} for the \textit{Janya} \textit{r\=aga}. We also analyse certain special cases of the \textit{Janya r\=aga} in order to understand their origins and structure better. 
%
\end{abstract}


\section{Introduction}
Carnatic music can be seen as a set of protocols coalesced with a musical scale, whose reference point is the fundamental frequency of the rendering vocalist/instrument. A musical scale, paired with it's unique set of protocols is called a \textit{r\=aga}. Due to the oral traditions of Carnatic music, the renditions differ for every performer. Moreover, a single \textit{r\=aga} may contain many different forms of compositions, making their standardisation a tough task. As a result, analysis of Carnatic music poses a lot of non-trivial challenges. In this article, we aim to establish causal relationships between different \textit{M\=e\d{l}akarta r\=aga} and their \textit{Janya} in Carnatic music. 

The research in Carnatic music analysis has been able to achieve \textit{r\=aga} classification \cite{madhusudhan2019deepsrgm}, using modern Transformer-based architectures. While \textit{r\=aga} classification is an interesting problem, the analysis of Carnatic music is not limited to that. Understanding the melody of a Carnatic composition \cite{koduri2011computational} is a non-trivial problem in itself. Other research problems include \textit{gamaka} and motif identification \cite{krishna2012carnatic} and tonic pitch identification \cite{salamon2012multipitch, gulati2014automatic, bellur2012knowledge}, where in all the articles show different approaches for the same problem. \cite{garani2019} shows a representation of Carnatic \textit{r\=aga} as Markov Chains. We have used a Markov chain representation inspired by the one introduced by Garani et. al. in \cite{garani2019}.

Causal analysis of music is generally studied for it's neurological and psychological effects. The Mozart Effect \cite{jenkins2001mozart, perlovsky2012mozart} is essentially the study of the cognitive function of the brain as a response to the usage of Mozart compositions as stimuli. \cite{van_Esch_2020} uses Granger Causality for inference of effective connectivity in brain networks for detecting the Mozart effect. 

\subsection{Carnatic Music: A Primer}

%
\textit{Svataha Ranjayitihi S\'varaha} is a popular Sanskrit phrase, which means that the \textit{s\'vara} sounds sweet even without embellishments. Basically, \textit{s\'vara}\footnote{\textit{S\'vara} and \textit{r\=aga} are both a disambiguation. Vernacular terms use the same words for both singular and plural forms.} is a melodious sound. The foundations of the Indian system of music are built upon the \textit{Sapta-s\'vara}, i.e. $\{Sa, Ri, Ga, Ma, Pa, Dha, Ni\}$, also denoted as $\{S, R, G, M, P, D, N\}$. Each note is independent and has its own position, known as the \textit{S\'varasthana}. 
 

The \textit{Abhinava R\=aga Manjari} defines the concept of \textit{\'sruti} as 
\begin{quote}
    \textit{N\=ityam G\=it\=opay\=ogitvam Bhigyeatvam Py\=atu, Lak\d{s}ae Proktam Supariy\=aptam Sa\.ng\=it \'Sritu Lak\d{s}a\d{n}a\.m}
\end{quote}
or, ``a certain distinctly identifiable sound that can be used in music is called \textit{\'sruti}''. Basically, \textit{\'sruti} refers to the position of the musical notes on the frequency spectrum, relative to the fundamental frequency. \textit{\'Sruti} is loosely analogous to the pitch of the sound. Usually, in Western music, the pitch of $A4 = 440~Hz$, also known as the concert pitch, is used as a reference point for tuning musical instruments \cite{Mueller15_FMP_SPRINGER}. In Indian classical music, the \textit{\'sruti} is tuned with reference to the tonic pitch of the lead singer/instrument. The central frequency of the tonic pitch is almost always more than an octave lower than $440~Hz$, and is different for every singer/instrument.

The concept of the \textit{r\=aga} is perhaps the crown jewel in the innovations of Indian classical music.
\textit{Brihaddesi} by M\=atanga Muni is the first text to define the concept of \textit{r\=aga}, according to which, it is a group of luminous notes with an integrated discipline of \textit{\'sruti} relationship, a power to steer the mind and evoke sentiment, and a built in capacity for infinite expansion \cite{ayyangar1972history}. 

Furthermore, a \textit{r\=aga} is required to have at least four \textit{s\'vara}. \textit{R\=aga} can broadly be classified into \textit{Janaka} or \textit{Janya}. \textit{Janaka} or the parent \textit{r\=aga} contain the \textit{Sapta-s\'vara} in both \textit{\=ar\=oha\d{n}a} and \textit{avar\=oha\d{n}a} i.e, ascending and the descending progressions respectively. 
\begin{equation}
\label{eq: aroha}
    \{S, R, G, M, P, D, N\}
\end{equation}

Equation \eqref{eq: aroha} shows the \textit{\=ar\=oha\d{n}a krama} of a \textit{Janaka} r\=aga. Equation \eqref{eq: avaroha} shows the \textit{avar\=oha\d{n}a krama} of the same \textit{r\=aga}.\footnote{Overdot on $S$ ($\dot{S}$) refers to the higher octave. Refer Section \ref{sec: auto_parse} for the details of the notations used.}

\begin{equation}
    \label{eq: avaroha}
    \{\dot{S}, N, D, P, M, G, R\}
\end{equation}

The \textit{s\'vara} in a \textit{Janaka r\=aga} must be present in a sequential order, as shown in Equation \eqref{eq: aroha} and \eqref{eq: avaroha}. The \textit{avaroha\d{n}a} must contain the same \textit{s\'vara} as the \textit{\=ar\=oha\d{n}a}. These are the essential qualities of a \textit{Janaka r\=aga}. 

\textit{Janya r\=aga}, or ``child'' \textit{r\=aga} are \textit{r\=aga} whose scale is derived from a \textit{M\=e\d{l}akarta} scale. These scale can be \textit{vakra} (zig-zag in pattern), or \textit{varja} (absence of certain notes).  

\textit{Janya r\=aga} are derivative \textit{r\=aga}, and if all the \textit{s\'vara} are taken from the parent \textit{r\=aga}, it is known as the \textit{Up\=anga Janya}. \textit{R\=aga Ha\d{s}adhwani} is an example of this kind of \textit{Janya r\=aga}. In case an \textit{anya s\'vara} (foreign note) is present it is known as a \textit{Bh\=a\d{s}\=anga r\=aga}. \textit{R\=aga K\=ambh\=oji} is an example of a \textit{Bh\=a\d{s}anga r\=aga}.    

A \textit{Bh\=a\d{s}\=anga r\=aga} can manifest in two ways. One, where the foreign note appears in the scale itself. \textit{R\=aga Bhairavi} scale contains the both $D_1$ and $D_2$, while it's \textit{Janaka r\=aga Natabhairavi} contains only $D_1$. So, in this case, $D_1$ is a \textit{sv\=akiya s\'vara} (note that belongs to the \textit{Janaka} scale), while $D_2$ is the \textit{anya sv\'ara}.  The other way of manifestation is where the \textit{anya s\'vara} appears only in rendition/performance. For example, \textit{r\=aga Bilahari}, \textit{Janya} of \textit{r\=aga Dh\=ira\'sankar\=abhara\d{n}a\.m} (commonly referred as \textit{{\'Sankar\=abhara\d{n}a\.m}}), employs $N_2$ in certain phrases during the rendition, whereas, $N_3$ is the \textit{sv\=akhya s\'vara}.

The \textit{Janaka}, which are also described as the \textit{M\=e\d{l}akarta r\=aga} have been devised using the \textit{Katapay\=adi Sa\.nkhy\=a} \cite{carnaticBook}  system, by Govindacharya in $17^{th}$ C. E. The 72 \textit{r\=aga} have been created/devised employing a twelve tonal scale, as shown in Table \ref{tab: twelve-tonal}.

The \textit{Janya} \textit{r\=aga} did not really originate from a \textit{M\=e\d{l}a}. Some of the \textit{Janya} \textit{r\=aga} are chronologically older than their \textit{M\=e\d{l}akarta} because the \textit{M\=e\d{l}akarta} were created after these \textit{Janya} \textit{r\=aga} were, similarly to how grammatical structures may have developed after the spoken language did. At that point, the \textit{M\=e\d{l}akarta} were named after the popular \textit{r\=aga} that were categorised under them.

\subsubsection{Compositional Structure}

Carnatic music is fundamentally based on compositions. Compositions are a reflection of a composer's musical prowess, and usually are structured to emphasize the true essence of a \textit{r\=aga}.  There are various compositional formats in Carnatic music, including, but not limited to \textit{g\=ita\d{m}}, \textit{s\'varajati}, \textit{vara\d{n}am}, \textit{kriti}, \textit{j\=ava\d{l}i}, \textit{padam} and \textit{till\=ana}. These vary in structure, size, style and features.

\textit{Kriti} is the most widely performed format in performances. It's structured in three main sections: the \textit{pallavi}, the \textit{anupallavi} and the \textit{cara\d{n}a}. Some derivatives of \textit{kriti} employ  \textit{sama\d{s}\d{t}hi cara\d{n}a} or \textit{ci\d{t}\d{t}ai s\'vara} as well. Likewise, \textit{var\d{n}am} begins with the \textit{pallavi}, followed by an \textit{anupallavi, ci\d{t}\d{t}ai s\'vara, cara\d{n}a} and \textit{cara\d{n}a s\'vara}.

\subsection{Causal Analysis}

Causal analysis is a study of correlation and directional influence between different variables (as measured using time-series or sequences). The effect that a measurable perturbation in one variable (time-series/ sequence) has on the another variable helps in evaluating the causal influence that exists between the two variables or systems under study (or in a network of interacting variables). According to Judea Pearl, a pioneer in the field of causal inference, causality can be pictured as a ladder with three rungs \cite{pearl2018book}. The lowest rung is called \textit{association}, which is the process of finding correlations between variables/time-series sequences. Quite often, two highly correlated time-series may not be causally related at all, or might share a mutual cause in a third completely different variable/time series. To solve these conundrums, the middle rung of the ladder of causality, called \textit{intervention} is used. Intervention is the process of finding the effect that a measurable perturbation of one variable has on another in a multi-variate system. The highest rung in the ladder of causality is \textit{counterfactuality}. In simple terms, counterfactual reasoning asks the question ``{\it what if}'' a particular intervention was carried out (hypothetically) - and aims to {\it imagine} its  repercussions or consequences to determine the directional causal linkage between two or more variables. 


For example, consider three variables producing time-series measurements: $T_t, T_s$ and $T_c$, where $T_t$ contains data about the atmospheric temperature of a region for the past few weeks, $T_s$ represents the trends in ice-cream sales and $T_c$ represents the trends in electricity consumption per day. Let's say a non-decreasing trend is observed in all of them. Association will tell us that each pair is highly correlated. Let's say a natural intervention occurs, and it rains for the next week. As a result, $T_t$ now becomes a decreasing sequence. If this results in both $T_s$ and $T_c$ having decreasing trends, then we can say that $T_t$ is the causal series, whereas $T_s$ and $T_t$ are effects. Counterfactual reasoning suggests considering alternate hypothetical situations like doubling the price of ice-creams or increasing the energy charges. In the former case, $T_s$ might get affected, but $T_t$ does not get affected. Moreover, a perturbation in $T_t$ will still affect $T_s$. Similarly, in the latter case, $T_c$ might get affected, but that does not affect $T_t$. In this scenario too, a perturbation in $T_t$ will still affect $T_c$. This verifies that $T_t$ is actually the causal series.     

The discovery of causal relationships between time-series sequences is useful in many scientific domains including but not limited to psychiatry, economics, neuroscience, climatology, medicine, physics and artificial intelligence. Refer \cite{kathpalia2021measuring} for a detailed report of these applications of causality.

\section{Methodology} 


Several sets of \emph{M\=e\d{l}akarta Janaka--\textit{Janya}} \textit{r\=aga} pairs were chosen for this analysis. The chosen \textit{r\=aga} are listed in Table \ref{tab: raga_data}.  The the number of compositions found in the corresponding \textit{r\=aga}, derived from \cite{shivkumarwebsite} has also been shown. In this article, we will represent a \textit{M\=e\d{l}akarta r\=aga} as $\mathcal{R}_{\eta}$
where $\eta$ stands for the \textit{M\=e\d{l}akarta} Index of that \textit{r\=aga} (also specified in the column \texttt{\textit{M\=e\d{l}akarta Sa\.nkhy\'a}} of Table \ref{tab: raga_data}). Similarly, a \textit{Janya r\=aga} will be represented as $\mathcal{R}_{\eta}^{(\alpha)}$, where $\eta$, as before, specifies the \textit{M\=e\d{l}akarta} Index of the corresponding \textit{M\=e\d{l}akarta r\=aga}, and $\alpha$ is the first letter of the \textit{r\=aga} name. For example, \textit{r\=aga Kharaharapriya}, which is the $22^{nd}$ \textit{M\=e\d{l}akarta r\=aga}, will be represented as $\mathcal{R}_{22}$, and \textit{r\=aga \=Abh\=ogi}, it's \textit{Janya}, will be represented as $\mathcal{R}_{22}^{(a)}$. 

$\mathcal{R}_{\eta}$ and $\mathcal{R}_{\eta}^{(\alpha)}$ are essentially sets that contain all the compositions in the corresponding \textit{r\=aga}. Hence, the cardinality $|\mathcal{R}_{\eta}|$ or $|\mathcal{R}_{\eta}^{(\alpha)}|$ equals the number of compositions in that set. This is what the \texttt{Num Comps} column in Table \ref{tab: raga_data} tells. Let $\textbf{C}$ be the set of all the compositions in the dataset. Then, $\textbf{C}$ is a super-set/container set of all the \textit{r\=aga}:
\begin{align}
    \label{eq: superset}
    \textbf{C} := \bigcup_{i \in \textbf{I}, \alpha \in \textbf{A}} \{\mathcal{R}_{i} \cup \mathcal{R}_{i}^{(\alpha)}\},
\end{align}
where $\textbf{I} := \{8, 15, 22, 28, 29, 65\}$ and $\textbf{A} := \{d, m, a, k, h\}$. Naturally, the cardinality $|\textbf{C}| = 57$.

The column \texttt{Identifier} of Table \ref{tab: raga_data} are just tags that were used to uniquely identify a \textit{r\=aga} in the code-base. Affirmatively, it is inspired by the $\mathcal{R}_{\eta}$/$\mathcal{R}_{\eta}^{(\alpha)}$ naming convention.

Notice there does not exist a specific \textit{Janya r\=aga} for \textit{r\=aga Mecakaly\=a\d{n}i} (commonly refered as \textit{Kaly\=a\d{n}i})  in Table \ref{tab: raga_data}. \textit{R\=aga Ha\.msadhwani} is considered a \textit{Janya} of \textit{r\=aga \'Sa\.nkar\=abhara\d{n}a\.m}, but, it can also be derived from \textit{r\=aga Kaly\=a\d{n}i}. \textit{R\=aga \'Sa\.nkar\=abhara\d{n}a\.m} employs \textit{\'Suddha Madhyama} ($M_1$) while \textit{r\=aga Kaly\=a\d{n}i} employs the \textit{Prat\=i Madhyama} ($M_2$). Notice that any of those \textit{Madhyama} are missisng from the \textit{r\=aga Ha\.msadhwani} scale. This paves the way for the aforementioned classification.  



\begin{table*}[!hbt]
    \caption{The list of all the \textit{r\=aga} used in this analysis. The symbol $\hookrightarrow$ represents the \textit{\=ar\=oha krama} while the symbol $\hookleftarrow$ represents the \textit{avar\=oha krama}.}
    \begin{tabular}{||>{\raggedleft}p{1.5cm}|c|l|P{1.5cm}|P{2cm}||}
    \hline \hline
        \textbf{Identifier} & \textbf{\textit{R\=aga}} & \textbf{Scale} & \textbf{Number of Compositions} & \textbf{\textit{M\=e\d{l}akarta Sa\.nkhy\'a}} \\
         \hline \hline
         \verb|8| & \textit{R\=aga Hanumat\=odi} & $\hookrightarrow$: $\{S, R_1, G_2, M_1, P, D_1,N_2, \dot{S}\}$ & $8$ & 8 \\
         & & $\hookleftarrow$: $\{\dot{S},N_2, D_1, P, M_1, G_2, R_1, S\}$ & &  \\\hline
         \verb|8_d| & \textit{R\=aga Dhanyasi} & $\hookrightarrow$: $\{S, G_2, M_1, P, N_2, \dot{S}\}$ & $5$ & 8 \\
         & & $\hookleftarrow$: $\{\dot{S},N_2, D_1, P, M_1, G_2, R_1, S\}$ & &  \\\hline 
         \verb|15| & \textit{R\=aga M\=ayama\b{l}avagau\b{l}a} & $\hookrightarrow$: $\{S, R_1, G_3, M_1, P, D_1, N_3, \dot{S}\}$ & $6$ & 15 \\
         & & $\hookleftarrow$: $\{\dot{S},N_3, D_1, P, M_1, G_3, R_1, S\}$ & &  \\\hline
         \verb|15_m| & \textit{R\=aga Malahari} & $\hookrightarrow$: $\{S, R_1, M_1, P, D_1, \dot{S}\}$ & $4$ & 15 \\
         & & $\hookleftarrow$: $\{\dot{S}, D_1, P, M_1, G_3, R_1, S\}$ & &  \\\hline
        \verb|22| & \textit{R\=aga Kharaharapriya} & $\hookrightarrow$: $\{S, R_2, G_2, M_1, P, D_2, N_2, \dot{S}\}$ & $6$ & 22 \\
         & & $\hookleftarrow$: $\{\dot{S}, N_2, D_2, P, M_1, G_2, R_2, S\}$ & &  \\\hline
         \verb|22_a| & \textit{R\=aga \=Abh\=ogi} & $\hookrightarrow$: $\{S, R_2, G_2, M_1, D_2, \dot{S}\}$ & $3$ & 22 \\
         & & $\hookleftarrow$: $\{\dot{S}, D_2,  M_1, G_2, R_2, S\}$ & &  \\\hline
         \verb|28| & \textit{R\=aga Harik\=ambh\=oji} & $\hookrightarrow$: $\{S, R_2, G_3, M_1, P, D_2, N_2, \dot{S}\}$ & $4$ & 28 \\
         & & $\hookleftarrow$: $\{\dot{S}, N_2, D_2, P, M_1, G_3, R_2, S\}$ & &  \\\hline
         \verb|28_k| & \textit{R\=aga K\=ambh\=oji} & $\hookrightarrow$: $\{S, R_2, G_3, M_1, P, D_2, \dot{S}\}$ & $5$ & 28 \\
         & & $\hookleftarrow$: $\{\dot{S}, N_2, D_2, P,  M_1, G_3, R_2, S\}$ & &  \\\hline
         \verb|29| & \textit{R\=aga Dh\=ira\'sankar\=abhara\d{n}a\.m} & $\hookrightarrow$: $\{S, R_2, G_3, M_1, P, D_2, N_3, \dot{S}\}$ & $6$ & 29 \\
         & & $\hookleftarrow$: $\{\dot{S}, N_3, D_2, P, M_1, G_3, R_2, S\}$ & &  \\\hline
         \verb|29_h| & \textit{R\=aga Ha\d{m}sadhwani} & $\hookrightarrow$: $\{S, R_2, G_3, P, N_3, \dot{S}\}$ & $4$ & 29 \\
         & & $\hookleftarrow$: $\{\dot{S}, N_2,  P,  G_3, R_2, S\}$ & &  \\\hline
         \verb|65| & \textit{R\=aga Mecakaly\=a\d{n}i} & $\hookrightarrow$: $\{S, R_2, G_3, M_2, P, D_2, N_3, \dot{S}\}$ & $6$ & 65 \\
         & & $\hookleftarrow$: $\{\dot{S}, N_3, D_2, P, M_2, G_3, R_2, S\}$ & &  \\\hline
    \end{tabular}
    \label{tab: raga_data}
\end{table*}

\subsection{Data Pre-processing}
\label{sec: auto_parse}
\begin{figure*}[!hbt]
    \centering
    \includegraphics[scale=0.4]{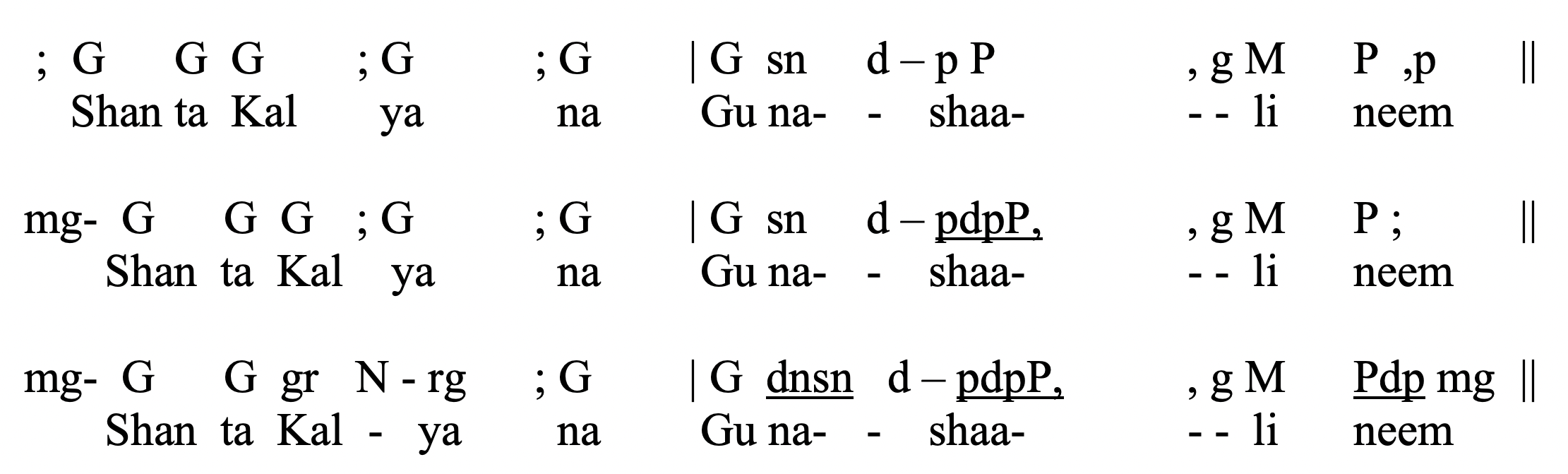}
    \caption{A typical Carnatic notation. This specific image is an excerpt from \textit{\'Sivak\=am\=e\'svari} by D\=ik\d{s}hit\=ar in the \textit{r\=aga} \textit{Kaly\=a\d{n}i}, taken from~\cite{shivkumarwebsite}.}
    \label{fig: representation}
\end{figure*}

Figure \ref{fig: representation} shows a typical way of annotating compositions of Carnatic music.  An automatic parser was developed to parse these notations into symbolic sequences.  The \textit{sv\'ara} $\{S, R, G, M, P, D, N\}$ or $\{s, r, g, m, p, d, n\}$ (the difference between the capital and small letters is that the capitals (e.g. $S$) occupy one count whereas the smalls (e.g. $s$) occupy half a count) were denoted by indices $\{0, 1, \ldots 6\}$ in a seven-tonal scale, and from the set $\{0, 1, 2, \ldots, 11\}$ if a twelve-tonal scale was used, depending upon the notes used in the \textit{r\=aga}. The rests ($;$ and $,$) are given the index $\infty$. Table \ref{tab: seven-tonal} shows the seven-tonal representation of the indices. Table \ref{tab: twelve-tonal} depicts notes used across Carnatic music and their corresponding indices. Hence, $S$ in the higher octave (usually denoted as $\dot{S}$) will be denoted by the Index $7$ in the seven-tonal scale, and by Index $12$ in a twelve-tonal one, $\dot{R}$ will be $8$ or $13/14/15$ respectively (latter depending upon the \textit{r\=aga}) and so on. Similarly, notes in the lower octave will have negative indices ($\underdot{N} \rightarrow -1/-2$ depending upon the \textit{r\=aga}).

\begin{table*}[h!]
    \caption{The seven-tonal scale, as has been used in the scope of this article. The actual note selected is decided by the scale of the (chosen) \textit{M\=e\d{l}akarta r\=aga}}
    \begin{tabular}{||l|c|c|c|c|c|c|c||}
    \hline \hline
    \textbf{Index} & $0$ & $1$ & $2$ & $3$ & $4$ & $5$ & $6$ \\ \hline
    \textbf{Symbol} & $S$ or $s$ & $R$ or $r$ & $G$ or $g$ & $M$ or $m$ & $P$ or $p$ & $D$ or $d$ & $N$ or $n$   \\  \hline \hline
    \end{tabular}
    \label{tab: seven-tonal}
\end{table*}


For example, consider \textit{r\=aga} \textit{M\=ayama\b{l}avagau\b{l}a}, whose "scale"\footnote{Both \textit{\=ar\=oha\d{n}a} and \textit{avar\=oha\d{n}a} \textit{krama} use the same notes.} can be defined with the notes  \\$\{S, R_1, G_3, M_1, P, D_1, N_3\}$. In a seven-tonal scale, it will simply be represented as $\{0, 1, 2, 3, 4, 5, 6\}$, since it is a \textit{samp\=ur\d{n}a M\=e\d{l}akarta r\=aga}. However, in a twelve-tonal scale, it will be represented as $\{0, 1, 4, 5, 7, 8, 11\}$; where the indexing follows  the scheme given in Table \ref{tab: twelve-tonal}. Now consider \textit{r\=aga Malahari}, the \textit{Janya} of \textit{r\=aga} \textit{M\=ayama\b{l}avagau\b{l}a}. As shown in Table \ref{tab: raga_data}, the \textit{\=ar\=oha} and \textit{avar\=oha} sequences are not equal. In a seven-tonal scale, the \textit{\=ar\=oha} will be $\{0, 1, 3, 4, 5\}$ and the \textit{avar\=oha} will be $\{0, 1, 2, 3, 4, 5\}$ (in the reverse order). In a twelve tonal scale, they will be $\{0, 1, 5, 7, 8\}$ and $\{0, 1, 4, 5, 7, 8\}$. For the purpose of parsing such a scale, we simply use the larger set amongst \textit{\=ar\=oha} and \textit{avar\=oha}. The absence of notes in the latter will affect the probability of occurrence of those notes in the \textit{r\=aga}'s compositions.

\begin{table*}[h!]
    \caption{The twelve-tonal scale in Carnatic music.}
    \begin{tabular}{||l|c|c|c|c|c|c|c|c|c|c|c|c||}
    \hline \hline
    \textbf{Index} & $0$ & $1$ & $2$ & $3$ & $4$ & $5$ & $6$ & $7$ & $8$ & $9$ & $10$ & $11$ \\ \hline
    \textbf{Symbol} & $S$ & $R_1$ & $R_2$ & $R_3$ &  & $M_1$ & $M_2$ & $P$ & $D_1$ & $D_2$ & $D_3$ &   \\ \hline
                    & & & $G_1$ & $G_2$ & $G_3$ & & & & & $N_1$ & $N_2$ & $N_3$ \\ \hline \hline
    \end{tabular}
    \label{tab: twelve-tonal}
\end{table*}

 
 The \textit{M\=e\d{l}akarta--Janya} pairs have been parsed in the twelve-tonal scale. For causal discovery method that we employ~\cite{pranay2021causal}, it does not matter as the \texttt{Lempel-Ziv} algorithm works with the total number of different symbols, and does not look at their magnitudes. However, it should be noted that a seven-tonal scale makes sense only when comparing \textit{M\=e\d{l}akarta}--\textit{Janya} \textit{r\=aga} pairs, wherein the scale of the latter is actually a subset of the former. Care must be taken that both \textit{M\=e\d{l}akarta} and it's \textit{Janya} have been parsed using the same scale. In short, as long as the compared sequences are parsed using the same scale, it works. 

As Figure \ref{fig: representation} depicts, a typical Carnatic notation consists of the \textit{s\'vara} used, the \textit{s\=ahitya} (the lyrics of the composition), and the corresponding meaning of the \textit{s\=ahitya}. Also, to guide the reader about the rendition, an attempt is made to place a copy of the \textit{s\=ahitya} in line with the \textit{svara} (Figure \ref{fig: representation} shows that). However, \textit{s\=ahitya}, their corresponding meanings and the rendition techniques are currently not adding any value to the input. Hence, only the \textit{s\'vara} part is considered as the input.

Another problem that we faced when analysing the notations obtained from \cite{shivkumarwebsite} was that the octave of a given note had to be inferred by the reader. The automatic identification of the octave of the note thus became a challenge. To get around this, we utilised the observation that note jumps with index difference more than seven semitones did not occur frequently in Carnatic music. Consider a note-event sequence belonging to some composition $\textbf{c} \in \textbf{C}$, and let $\|\textbf{c}\|$ denote the length of the sequence. Let $i \in [2:\|\textbf{c}\|-1]$. Now, if $|c_i - c_{i - 1}| > 7$ semitones\footnote{Seven semitones usually corresponds to five indices in the seven-tonal scale.}, then it is highly likely that $c_i$ was changed to the lower or the higher octave, depending upon the index of the note $c_i$. For $c_i \geq D$ (pitches higher than $D$), $c_i = c_i - x$ and for pitches lower than $D$, $c_i = c_i + x$, when $x$-tonal scale is used, $x \in \{7, 12\}$. In case $c_i = \infty$ (i.e., it is a rest), this comparison with $c_{i-1}$ was not performed. In case $c_{i-1} = \infty$, the $c_i$ was compared with $c_{i-2}$ and so on. In essence, the fact that a typical Carnatic composition spans two octaves $[\underdot{P}, \dot{P}]$ is harnessed. So, when a high-index note (such as $D$ or $N$) is immediately followed by a low-index note (such as $S$ or $R$), it usually means that either the former is in a lower octave or the latter is in a higher octave. This led to inconsistencies in the parsed compositions, but it was still used as it tracked the octave changes properly. 

Each composition is structured according to a rhythmic cycle, known as the \textit{t\=a\d{l}a}. Few popular Carnatic \textit{t\=a\d{l}a} are {\=Adi} (eight-beat rhythmic cycle), \textit{Kha\d{n}\d{d}a C\=apu} (five-beat rhythmic cycle), \textit{R\=upaka} (six-beat rhythmic cycle) and likewise. \textit{\=Avartana} in Carnatic music, is one whole rhythmic cycle within a \textit{t\=a\d{l}a} (e.g. completion of eight beats in \textit{\=Adi t\=a\d{l}a} marks the completion of an \textit{\=avartana}.) The use of the \textit{da\d{n}\d{d}\=a} ($|$) indicates the separation between the different divisions of the \textit{\=avartana}. The use of the ``double \textit{da\d{n}\d{d}\=a}'' ($||$) (\cite{carnaticBook}) in the notations marks the end of the \textit{\=avartana} of the \textit{t\=a\d{l}a} being used. In the parser, only the double \textit{da\d{n}\d{d}\=a} is used to track the number of measures encountered thus far.

The issue of octave resolution also makes it rather tough to distinguish between the \textit{\=ar\=oha} phrases and the \textit{avar\=oha} phrases. Consequently, for the final parse, a vocabulary built from the union of all the notes in the \textit{\=ar\=oha} and the \textit{avar\=oha} scales is ultimately used. For \textit{M\=e\d{l}akarta r\=aga}, this is not a problem, as both the \textit{\=ar\=oha krama} and the \textit{avar\=oha krama} use all the seven \textit{s\'vara}. In case of \textit{Janya r\=aga}, some \textit{s\'vara} might be missing from the \textit{\=ar\=oha}, while other might be missing from the \textit{avar\=oha}. Redundant \textit{s\'vara}, which may occur in some \textit{vakra prayoga}, are eliminated. In this case, a union of both is used as the vocabulary. For example, consider \textit{r\=aga Malahari} from Table \ref{tab: raga_data}. The vocabulary of \textit{r\=aga Malahari} will look like
\begin{align}
    \label{eq: scale}
    \{S, R_1, G_3, M_1, P, D_1\}.
\end{align}

The parsed notations are represented in three dimensions: the frequency space (the note index), the length space (relative note duration/number of beats per note) and the temporal space (the measure indices). Every note-event can thus be represented as 
\begin{align}
    \label{eq: 3d-space}
    \Vec{n} \in \mathbf{S} \iff \Vec{n} &:= n_a \hat{a} + n_b \hat{b} + n_c \hat{c},
\end{align}
where $\hat{a}$, $\hat{b}$ and $\hat{c}$ are unit vectors in the directions of note-indices, number of counts and measure indices respectively. $(\mathbf{S})$ is a discrete three-dimensional vector-space  with basis vectors $\hat{a}$, $\hat{b}$ and $\hat{c}$, capable of representing any song/composition from parsed notations.

It is quite likely that the Carnatic parser will encounter the occurrence of more counts than the \textit{\=avartana} count. For example, try counting the number of counts in each \textit{\=avatanam} in Fig \ref{fig: representation}. One time out of two, this is the case. This is because the underlined notations (e.g. $\underline{dnsn}$) or even notations without spaces (e.g. $mg-$) are supposed to be counted as $1$ count. However, the parser is not developed enough to handle this. As of now, every symbol is given it's designated duration, and if the total exceeds the \textit{\=avartana} counts, a normalization is done. Let $\textbf{n} := (\Vec{n_1}, \Vec{n_2}, \ldots, \Vec{n_k})$ represent a complete \textit{\=avartana} for some composition $c \in \textbf{C}$. Let $\theta$ be the number of beats per \textit{\=avartana}. If $\sum_{i=1}^k (n_i)_b > \theta$, then:
\begin{align}
    \label{eq: beat-norm}
    (n_i)_b' = \frac{(n_i)_b}{\sum_{i=1}^k (n_i)_b} \times \theta ~~~ \forall i \in [1:k]
\end{align}
where $(n_i)_b'$ denotes the new $\hat{b}$ component in every $\Vec{n_i}$.

For example, consider the phrase $G ~ \underline{dsns} ~ d ~\underline{pdpP,} ~ g ~ M ~ P~;||$. The specified \textit{t\=ala} is \textit{\=Adi}, so this should be $\theta = 8$ counts. However, the total is $1 + 0.5*4 + 0.5 + (0.5*4+1) + 0.5 + 1 + 1 + 1 = 10$. Then the original duration of every event in the sequence is divided by $10$ and then multiplied by $8$, so the new durations are $0.8 + 0.4*4 + 0.4 + (0.4*4+0.8) + 0.4 + 0.8 + 0.8 + 0.8  = 8$.

As a result, the parsed duration may not be equal to the notated duration. However, even in real world, the note duration is adjusted through the rendition, and hence differs from the notated duration quite often.

\subsubsection{Pre-processing for \textit{R\=aga K\=ambh\=oji}}
\label{sec: kambhoji}

\textit{R\=aga K\=ambh\=oji} ($\mathcal{R}_{28}^{(k)}$) is a prime example of a \textit{bh\=a\d{s}\=a\.nga r\=aga}, wherein, a foreign note appears in the scale itself. As established earlier, it is hard to distinguish between an \textit{\=ar\=oha} and an \textit{avar\=oha} phrase. In \textit{r\=aga K\=ambh\=oji}, the foreign \textit{s\'vara} appears only as an extension of the \textit{avar\=oha}. Even if we add $N_3$ to the vocabulary, we need a way of differentiating between the two $Ni$-s and deciphering them in the notations. 

Fortunately, $N_3$ occurs only in the \textit{vakra prayoga}, which is a variation of $PND$ (e.g. $pnd$, $p;nd$, $p,n;,d$ etc). So, each time the parser encounters the \textit{s\'vara} $D$ or $d$ in a \textit{r\=aga K\=ambh\=oji} composition, two of the previously encountered \textit{s\'vara} are checked. In case the current $D$ (or $d$) follows a $N$ (or $n$), which is preceded by a $P$ (or $p$), the $N$ (or $n$) is the foreign $N_3$. This specific implementation was realized because the parser has knowledge of the past: i.e. it remembers all the notes it has encountered in the measure so far, but, it lacks the knowledge of the future. When we encounter a $N$, we cannot check whether the next note in the sequence  is $D$ or not. 

\subsection{Data Processing}
\label{sec: data_proc}

We can emulate the melody of the song by mapping the duration ($\hat{b}$) of every note-event to an integer and repeating that note that many times. Let \textbf{N} be the note-event sequence of a composition  in $\textbf{C}$, such that $\textbf{N} := (\Vec{n_1}, \Vec{n_2}, \ldots, \Vec{n_K})$, where each $\Vec{n_i} \in \mathbf{S}$. Then we define a mapping $f: \mathbf{S} \rightarrow \mathbf{S}$ such that: 
\begin{align}
    f(\Vec{n}) &:= (n_a + 36) \hat{a}  + \lceil n_b \cdot 480 \rceil \hat{b} + n_c \hat{c}.
\end{align}

Observe that the duration of every note-event gets multiplied by $480$. This was done because, commonly, the number of beats every note-event lasts for will be an inverse power of two, and in some cases, might be as small as $\frac{1}{32}$. However, other beat-lengths like \textbf{triplets} (that last for $\frac{1}{3}$ of a beat) or \textbf{fifth-beat} (that lasts for $\frac{1}{5}$ of a beat) are also encountered frequently. By multiplying the note-event duration by $480 = 32 \times 3 \times 5$, we hope to convert almost all the $n_b$'s to an integer. We might also encounter some special case scenarios (e.g., seventh-beats or derivatives of triplets and fifth-beats) in which case, just multiplying by $480$ would not convert $n_b$ to an integer. That is why the ceiling operation ($\lceil \cdot \rceil$) was used. 

If we now repeat each $f(\Vec{n}) \cdot \hat{a}$ (which represents a note index) $f(\Vec{n}) \cdot \hat{b}$ times (which is now an integer), we then obtain a new sequence $N^\dagger$ which emulates the melody of the song. For example, consider three random note events $\Vec{n}_1, \Vec{n}_2, \Vec{n}_3$. Let $\Vec{n}_1 = 4\hat{a} + 0.5\hat{b} + 1\hat{c}$, $\Vec{n}_2 = 5\hat{a} + 1 \hat{b} + 5 \hat{c}$ and $\Vec{n}_3 = 0\hat{a} + 0\hat{b} + 2\hat{c}$. Then, with the contribution of $\Vec{n}_1, \Vec{n}_2, \Vec{n}_3$, $N^\dagger = \{ 0, 0, 0\ldots(480 \text{ times}), 4, 4, 4\ldots(240, \text{ times}),\ldots,\\ 5, 5, 5\ldots(480 \text{ times}), \ldots \}$. Equation \eqref{eq: n_dagger} shows how the final song might be represented. The song has been transposed by three octaves (by adding $36$ to $n_a$) because the causal discovery algorithm  (implemented in the \texttt{ETCPy} library) does not support sequences with integers less than zero.


\begin{align}
    \label{eq: n_dagger}
    \mathbf{N}^\dagger = \{&\underbrace{f(\vec{n_1})_a, f(\vec{n_1})_a, \ldots}_{f(\vec{n_1})_b \text{ times}}, \underbrace{f(\vec{n_2})_a, f(\vec{n_2})_a, \ldots}_{f(\vec{n_2})_b \text{ times}}, \ldots,
    \underbrace{f(\vec{n_K})_a, f(\vec{n_K})_a, \ldots}_{f(\vec{n_K})_b \text{ times}} \}
\end{align}

\subsection{Causal Inference}
As highlighted earlier, the \texttt{ETCPy} library\footnote{Available on \url{https://github.com/pranaysy/ETCPy}, open source, with the Apache 2.0 licence.} has been used to compute the Causality between pairs of sequences. $N^\dagger$ for a particular composition has variable lengths. Lowest $N^\dagger$ values for each of the chosen \textit{M\=ed{l}akarta--Janya} are depicted in Table \ref{tab: surr_test_results}. 

\begin{figure*}[!hbt]
    \centering
    \includegraphics[width=\textwidth]{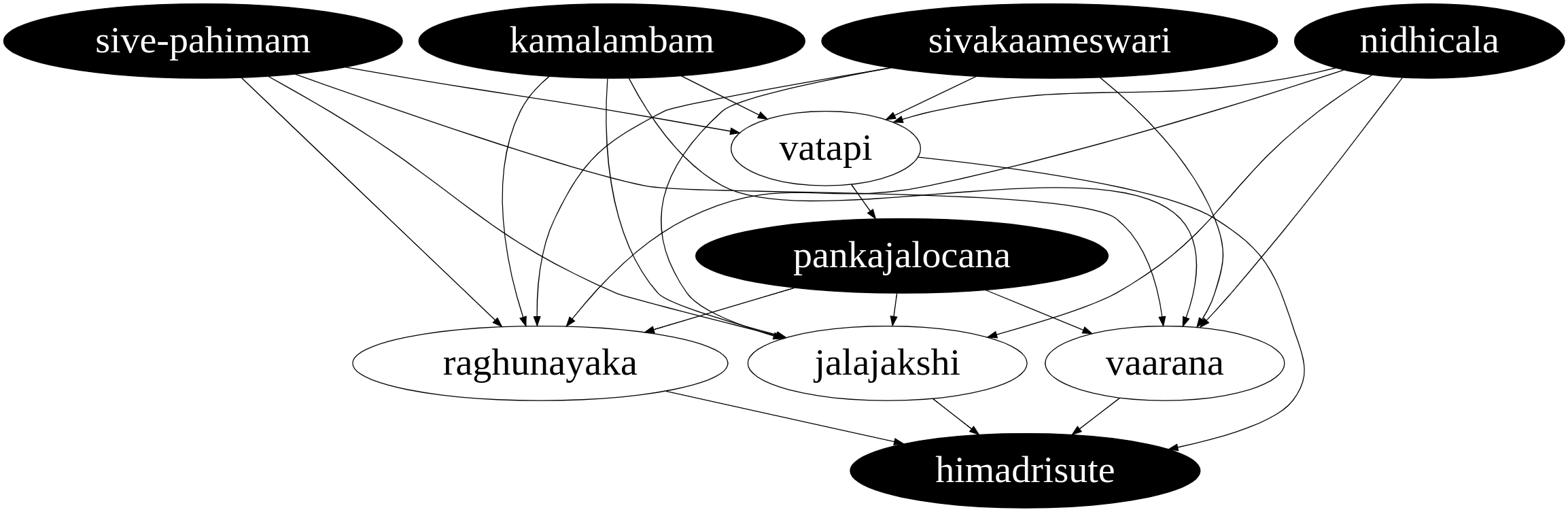}
    \caption{Results of LZP causal analysis on $\{\mathcal{R}_{65} \cup \mathcal{R}_{29}^{(h)}\}$. The black background highlights the \textit{M\=e\d{l}akarta} compositions.}
    \label{fig: caus_graph}
\end{figure*}



%
The \texttt{Lempel-Ziv-Penalty} (LZP for short, please see Appendix 1 for a detailed description) model has been used; which implies the algorithm used for compressing the sequences is \texttt{LZ-76} \cite{lempel1976complexity}. We then find the causal direction between each pair and output them as a Directed Acyclic Graph (DAG). The graph in Figure \ref{fig: caus_graph} depicts one such DAG. Precisely, it features a result with \textit{r\=aga Kaly\=a\d{n}i} as the \textit{M\=e\d{l}akarta} and \textit{r\=aga Ha\d{m}sadhwani} as its \textit{Janya}. It has a top-down structure. Hence, a node never connects to any other node above it, but may connect with multiple nodes below it. Moreover the black shade represents compositions in the \textit{M\=e\d{l}akarta} \textit{r\=aga}.


\subsection{Surrogate Testing}
\label{sec: surr_test}
\subsubsection{Surrogate Data Generation}
Following the parsing step highlighted in Section \ref{sec: auto_parse}, we end up with a data stream that can be completely represented in the vector space $\textbf{S}$. This data-stream is fairly accurate in the $\hat{b}$ dimension, not so accurate in the $\hat{a}$ dimension and exact in the $\hat{c}$ dimension. The parser ends up covering $(\underdot{S}, \Ddot{S})$ in compositions, using around $[19, 23]$ symbols as vocabulary in the $\hat{a}$ (note-index) dimension of a single \textit{r\=aga}. As \cite{garani2019} shows, a Carnatic \textit{r\=aga} can be represented as a layered graph. However, due to inconsistencies in the $\hat{b}$ (duration) dimension, we cannot follow their exact approach. 

\begin{figure*}[t]
\begin{subfigure}[b]{0.22\textwidth}
    \centering
    \includegraphics[width=\columnwidth]{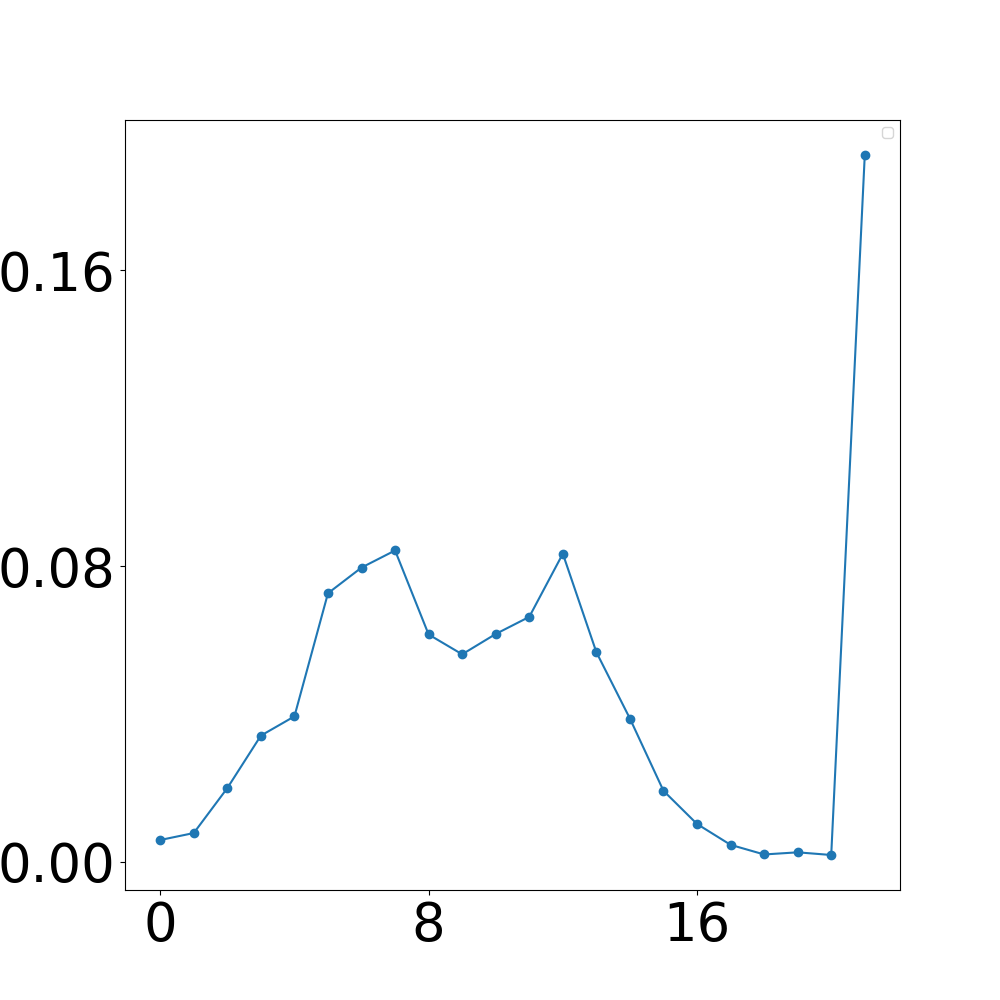}
    \caption{$\pi(\mathcal{R}_8)$}
    \label{fig: stationary_8}
\end{subfigure}
\hspace{0.02\textwidth}
\begin{subfigure}[b]{0.22\textwidth}
    \centering
    \includegraphics[width=\columnwidth]{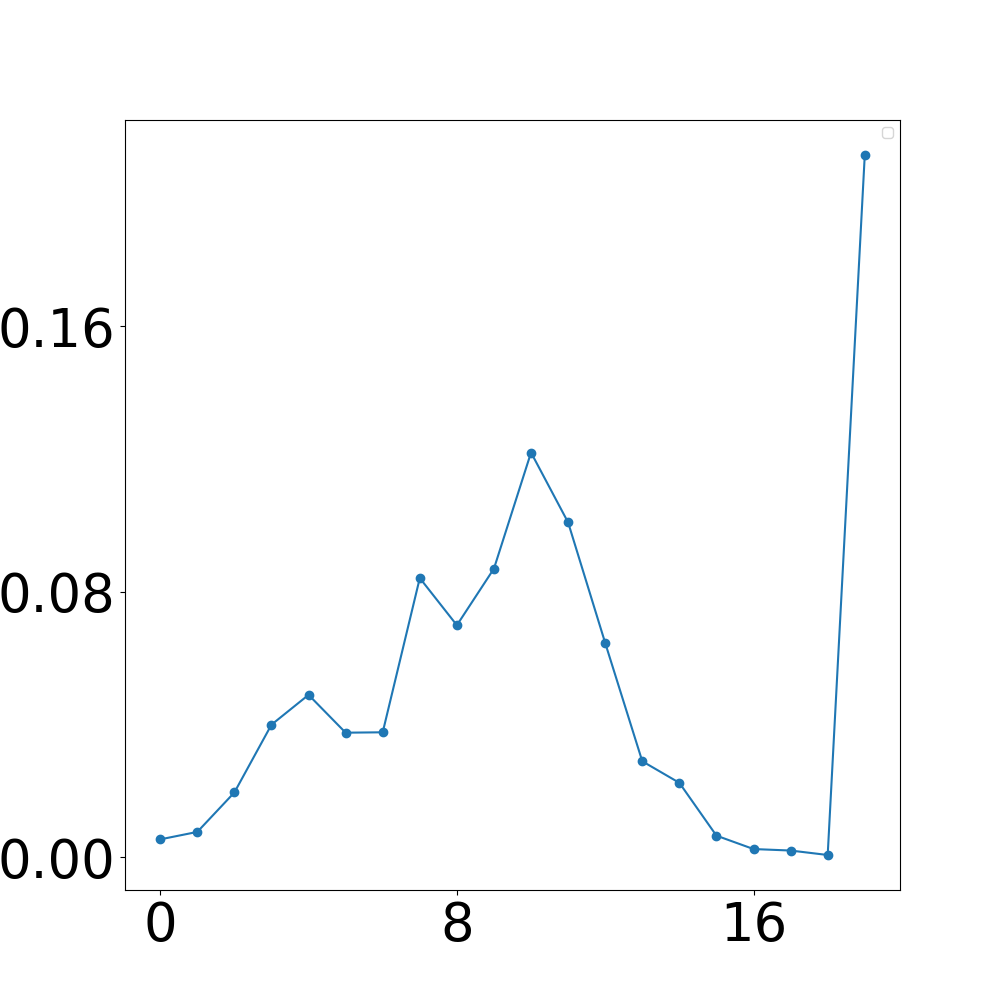}
    \caption{$\pi(\mathcal{R}_{15})$}
    \label{fig: stationary_15}
\end{subfigure}
\hspace{0.03\textwidth}
\begin{subfigure}[b]{0.22\textwidth}
    \centering
    \includegraphics[width=\columnwidth]{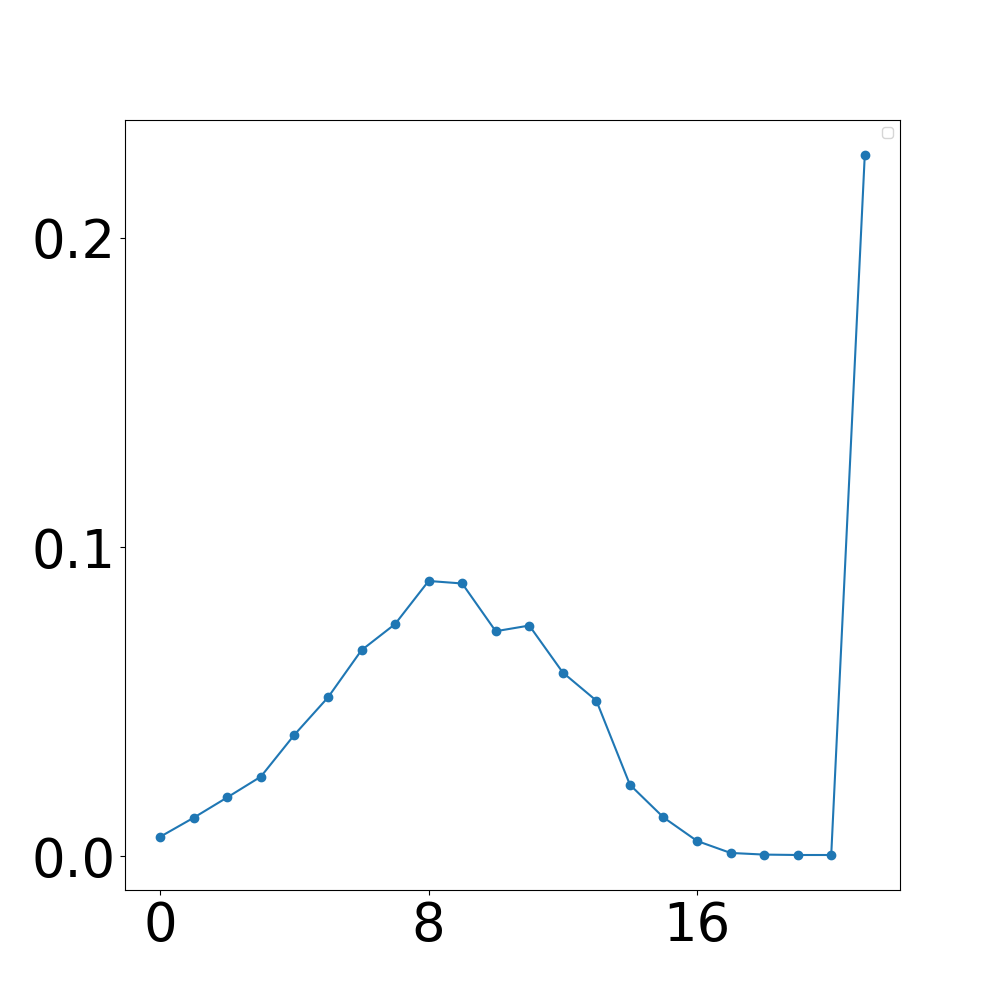}
    \caption{$\pi(\mathcal{R}_{22})$}
    \label{fig: stationary_22}
\end{subfigure}
\hspace{0.03\textwidth}
\begin{subfigure}[b]{0.22\textwidth}
    \centering
    \includegraphics[width=\columnwidth]{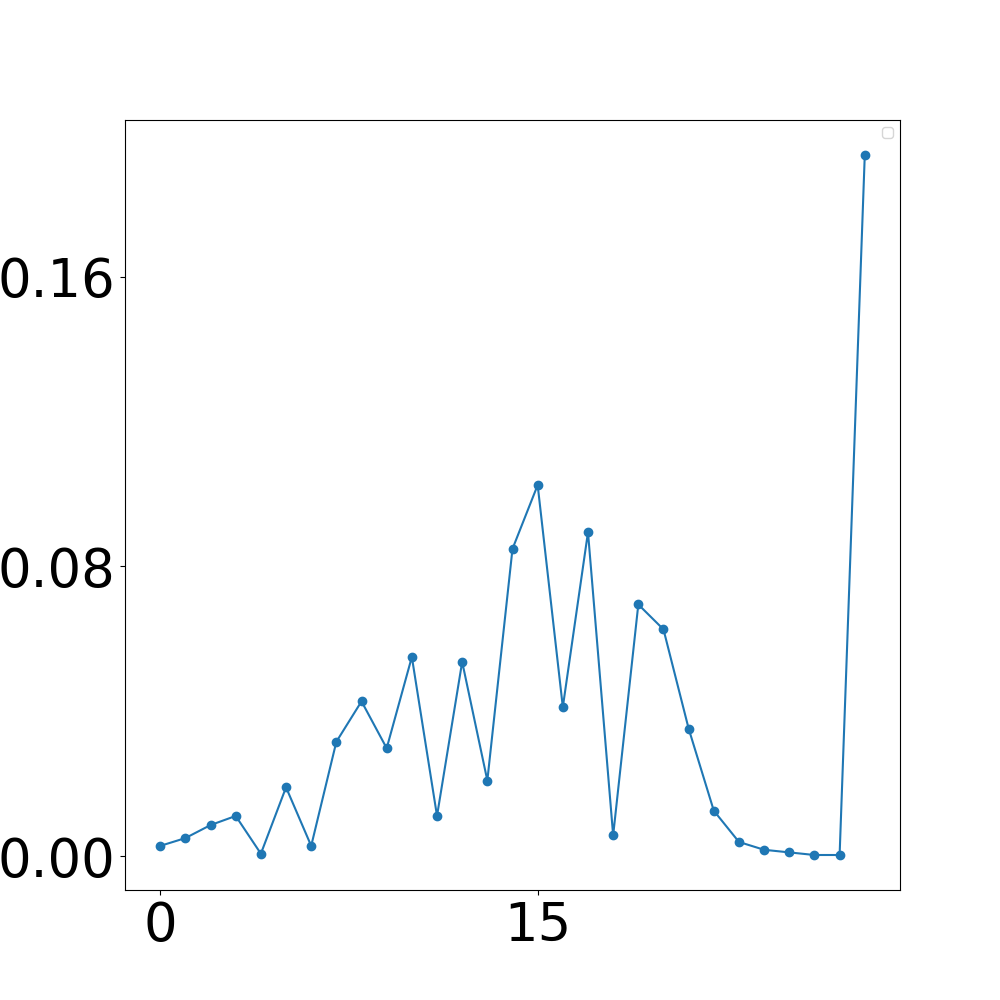}
    \caption{$\pi(\mathcal{R}_{28})$}
    \label{fig: stationary_28}
\end{subfigure}
\caption{The stationary distribution ($\pi$) for several \textit{r\=aga}. Note the singular point of high probability towards the end, which represents the rest. The other points indicate notes along approximately three octaves. The peaks are usually found in the central octave. The X-axis zero corresponds to the lowest pitch encountered in the \textit{r\=aga}.}
\label{fig: some_stationary_dists}
\end{figure*}

Hence, using just the $\hat{a}$ (note-index) dimension, we formulate Markov Chains (MC) for each \textit{r\=aga}. The stationary distribution $\pi$ (calculated using the transition matrix of an order-1 MC) for several \textit{r\=aga} is shown in Fig \ref{fig: some_stationary_dists}. As we are not using the $\hat{b}$ (duration) and $\hat{c}$ (measure-index) dimensions for this, we coalesce all the compositions in $\mathcal{R}_{\eta}$ or $\mathcal{R}_{\eta}^{(\alpha)}$ to form one single sequence $\textbf{R} = (\Vec{r_1}, \Vec{r_2}, \ldots, \Vec{r_{\gamma}})$, where $\Vec{r_i} \in \textbf{S}$. This way, we are able to define the \textit{r\=aga}'s statistics. To save memory space, the transition probabilities are found as-is; i.e. the transition is added to the matrix only if it is not already there, or if it is, its frequency is updated. For example, consider an order-3 MC. One possible transition is $P\underdot{D}\dot{S} \rightarrow \dot{R}$. Even if it is quite likely that $p(\dot{R}|\dot{S}) >0$, the occurrence of the phrase $P\underdot{D}\dot{S}$ itself is very unlikely. As it never occurs, it is never added to the transition matrix ($\rho$). Consequently, we avoid a sparse transition matrix. 

After calculating the required transition matrix using $\textbf{R}$, new compositions $\Phi := \{\phi_1, \phi_2, \ldots \phi_s\}$ (where $s$ is the number of surrogates) are generated, primarily using the transition probability matrix. Firstly, a ``key'' is uniformly selected at random from rows of $\rho$. In case of an order-1 MC, $\rho$ ends up being a square matrix. Also, in case of an order-1 MC:
\begin{align}
    \label{eq: stationary_dist}
    \pi = \pi \rho.
\end{align}
So, instead of using the uniform distribution, the stationary distribution ($\pi$) of the MC is used for key selection. Once the key $\kappa$ is selected, the next element is generated using $\rho[\kappa]$ as a weighted distribution. Let $\Vec{n_1}$ be the first element of some $\phi$. Then, $(n_1)_a \sim \rho[\kappa]$. After a valid $(n_1)_a$ generation, it is added to the end of $\kappa$ and one element is removed from the front of $\kappa$. Naturally, in case of order-1 MC, it is simply $\kappa = (n_1)_a$. 

The problem with order-2 or higher order MC's is that after several iterations, the generation seems to get stuck in an infinite loop of a single note. It is very likely that, for example, after several samples, all the new generations will be rests $;$. That is because, $p(;|;;) > 0$ or maybe even $p(;|;;;;) > 0$, while $p(\text{any other element}|;;;;) \approx 0$. Consequently, order-1 MC's tend to give the best results.

\begin{figure*}[!hbt]
\begin{subfigure}[b]{0.25\textwidth}
    \centering
    \includegraphics[width=\columnwidth]{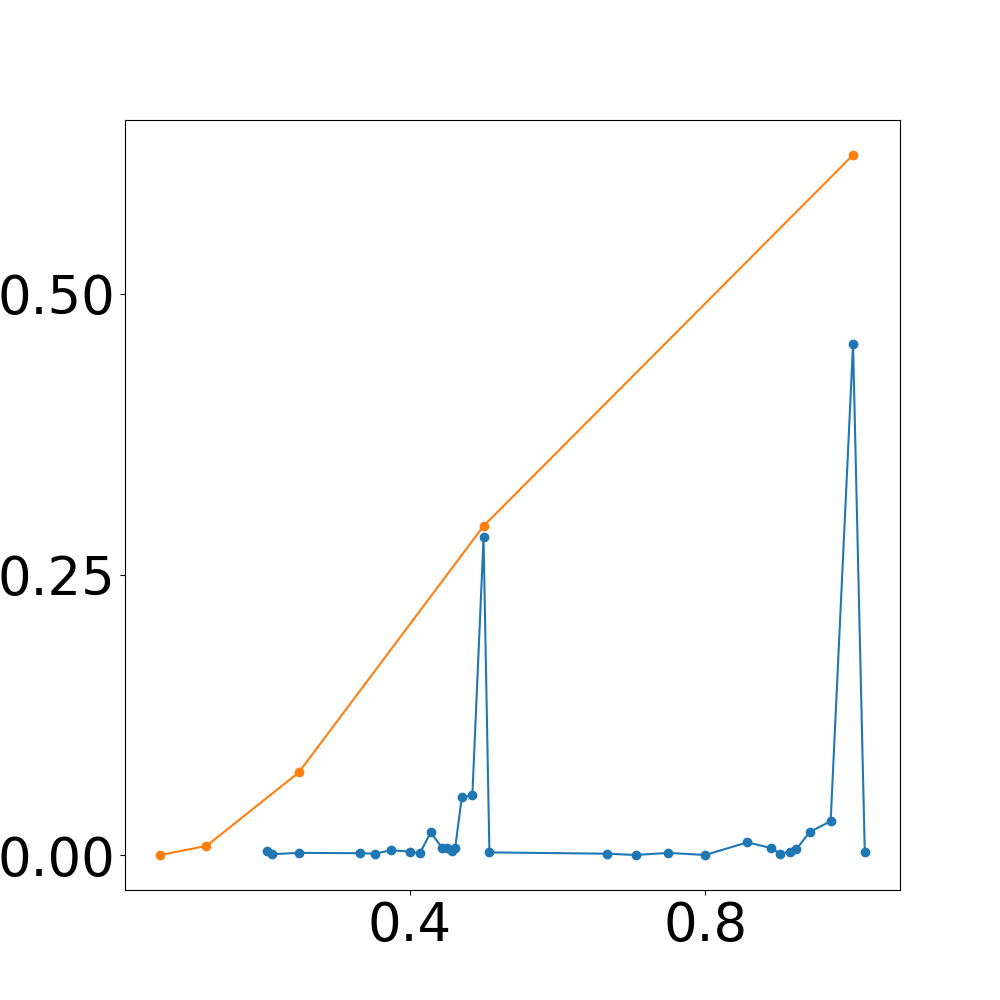}
    \caption{Overall frequency distribution of note-event durations for $\mathcal{R}_8$ (blue) and $\Phi(\mathcal{R}_{8})$ (orange).}
    \label{fig: dur_dist_8}
\end{subfigure}
\hspace{0.1\textwidth}
\begin{subfigure}[b]{0.25\textwidth}
    \centering
    \includegraphics[width=\columnwidth]{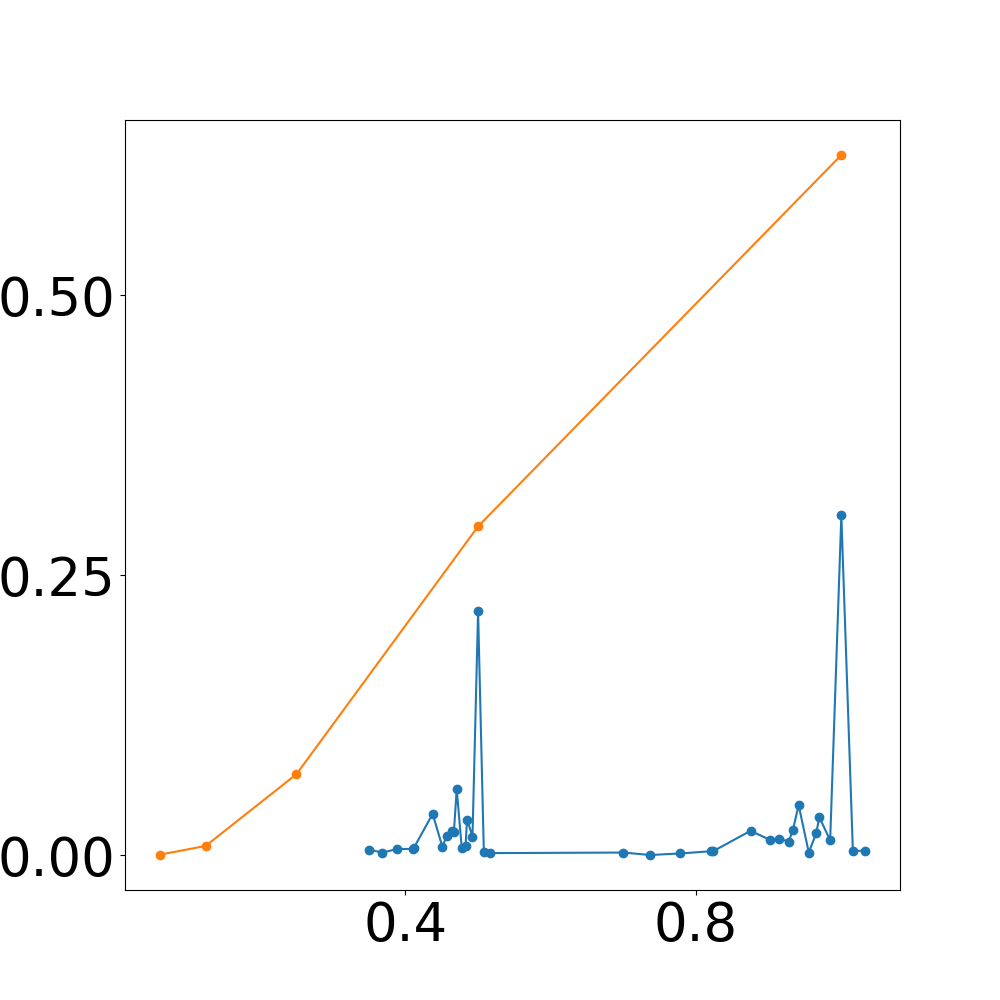}
    \caption{Overall frequency distribution of note-event durations for $\mathcal{R}_{28}^{(k)}$ (blue) and $\Phi(\mathcal{R}_{28}^{(k)})$ (orange).}
    \label{fig: dur_dist_28_k}
\end{subfigure}
\hspace{0.1\textwidth}
\begin{subfigure}[b]{0.25\textwidth}
    \centering
    \includegraphics[width=\columnwidth]{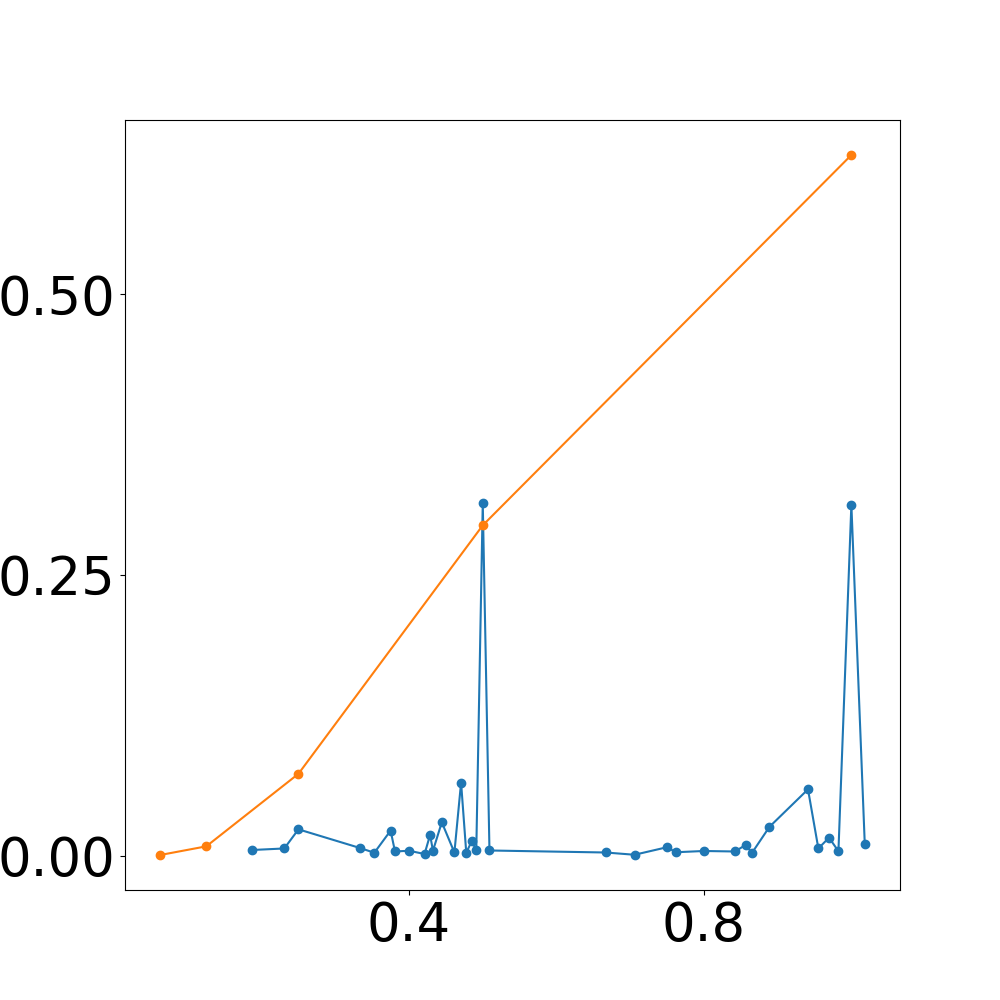}
    \caption{Overall frequency distribution of note-event durations for $\mathcal{R}_{15}$ (blue) and $\Phi(\mathcal{R}_{15})$ (orange).}
    \label{fig: dur_dist_15}
\end{subfigure}
\caption{Frequency distributions of note-event durations obtained from the parser (blue) and from the surrogate generations (orange). Even though they do not follow the same distributions, notice that the blue ones peak at the inverse powers of 2.}
\label{fig: dur_dists}
\end{figure*}

As explained in Section \ref{sec: auto_parse}, the event durations obtained by the parser are not necessarily accurate. So, we define a new distribution: 
\begin{align}
    \label{eq: dur_dist}
    \mathbbm{2} := 2^{-\lfloor \mathcal{N}^+(2, 1) \rfloor}
\end{align}
where $\mathcal{N}^+(2, 1)$ stands for the positive part of the normal distribution $\mathcal{N}(2, 1)$. So, $(n_1)_b \sim \mathbbm{2}$. As shown in Figure \ref{fig: dur_dists}, Equation \eqref{eq: dur_dist} gives out the values at which $(n_i)_b$ values peak.

The \textit{\=avartana} ($\tau$) count for $\phi$ is chosen uniformly at random from the set $\{6, 7, 8, 10, 12, 14, 16\}$. After several valid generations, if $|\sum_{i=t_1}^{t_2} (n_i)_b - \tau| <= \epsilon$, then $(n_{t_2})_c = (n_{t_2 - 1})_c + 1$, where $(n_{t_1})_c = (n_{t_1 - 1})_c + 1$ (i.e. the last measure). $\epsilon$ is a small constant, which has been given the value $0.0078125 = 2^{-7}$ for efficient generation. This makes $\mathbbm{2}$ the limiting distribution; i.e. $\mathbbm{2}$ is the only distribution that might reject a generated sample, given it does not meet the aforementioned conditions. Sometimes, the generated sample $g \sim \mathbbm{2}$ is too high to fit in the \textit{\=avartana} counts. In that case, the entire note event $\vec{n}$ is deleted, and new values are obtained for $\vec{n}_a$ and $\vec{n}_b$.  Initially, for $\Vec{n_1}$, $(n_1)_c = 0$.

In this manner, a thousand note events are generated for every $\phi$. The number of measures $\max n_c ~ \forall \Vec{n} \in \phi$ gets adjusted according to selected \textit{\=avartana} count. Fig \ref{fig: surr_statdists} shows the stationary distribution of fifty surrogates as compared to the original stationary distribution of \textit{r\=aga }. 

\begin{figure*}[!hbt]
\begin{subfigure}[b]{0.48\textwidth}
    \centering
    \includegraphics[width=0.5\columnwidth]{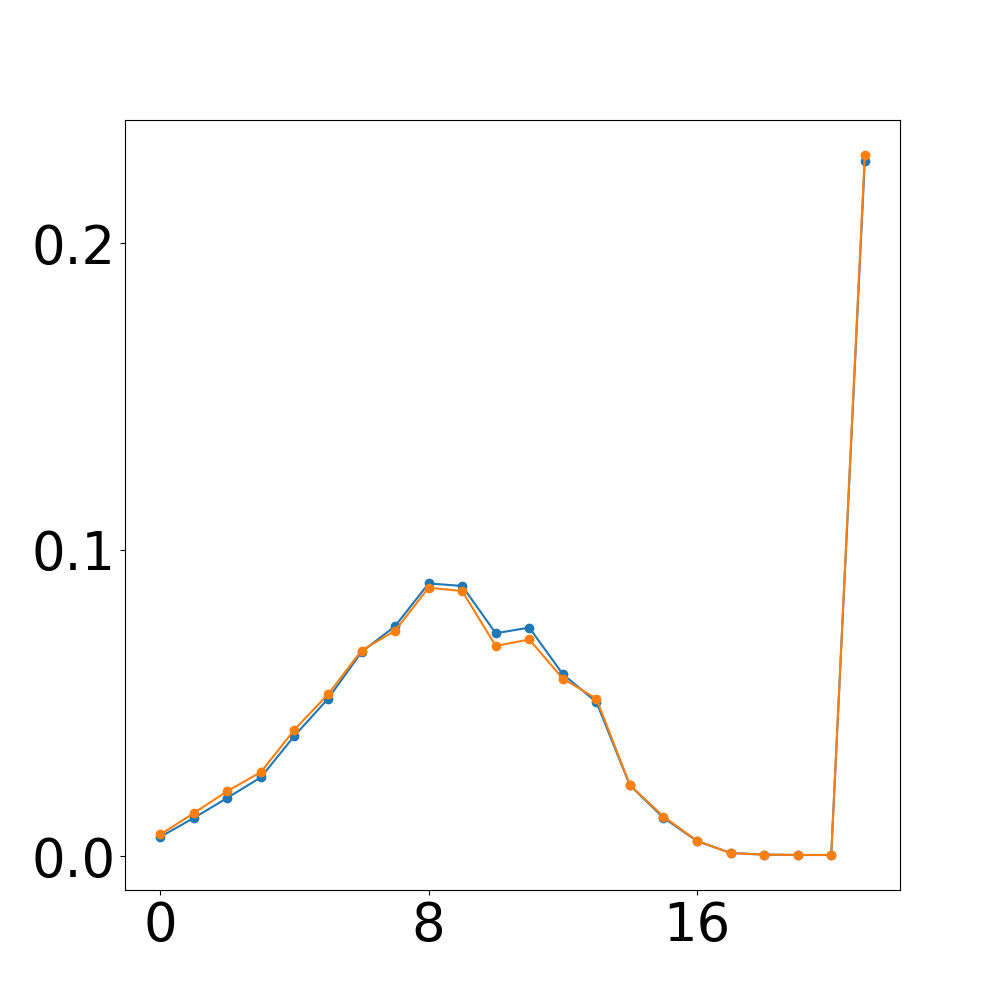}
    \caption{$\pi(\mathcal{R}_{22})$ (blue) and $\pi(\Phi(\mathcal{R}_{22}))$ (orange). }
    \label{fig: surr_stat_22}
\end{subfigure}
\begin{subfigure}[b]{0.48\textwidth}
    \centering
    \includegraphics[width=0.5\columnwidth]{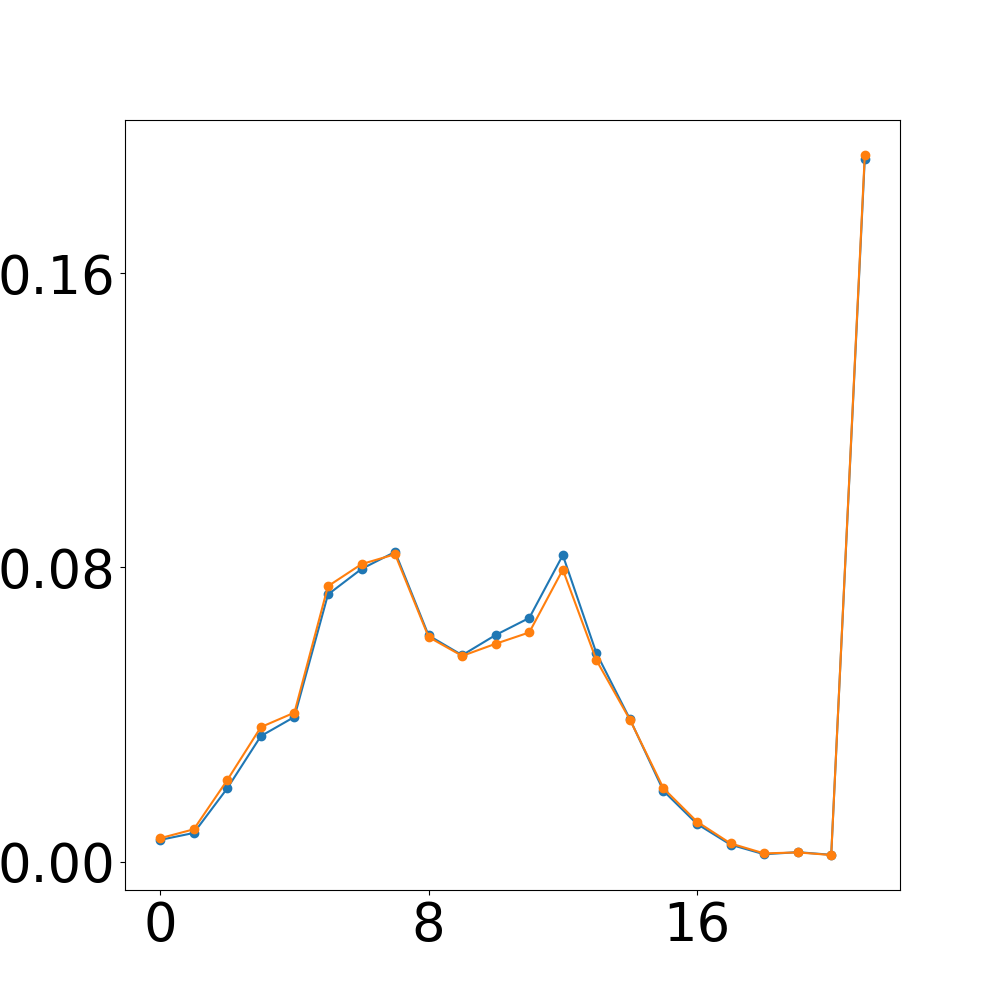}
    \caption{$\pi(\mathcal{R}_{8})$ (blue) and $\pi(\Phi(\mathcal{R}_{8}))$ (orange). }
    \label{fig: surr_stat_8_d}
\end{subfigure}
\caption{The stationary distribution of the specified \textit{r\=aga} compared to the stationary distribution of its surrogates.}
\label{fig: surr_statdists}
\end{figure*}

\subsubsection{Data Processing for Causal Analysis with Surrogate Data}
\label{sec: surr_data_proc}
 The complete sequence (as shown in Equation \eqref{eq: n_dagger}) is used only if its length is the minimum throughout the set. Otherwise, a random sub-sequence of the minimum length is chosen from the given sequence. e.g. let $\textbf{c}_i ~ \in ~ \{\mathcal{R}_{29} \cup \mathcal{R}_{29}^{(h)}\}$, where $i \in [1:10]$. Let $N_{min} = \min(\|N^{\dagger}_1\|, \|N^{\dagger}_2\|, \ldots, \|N^{\dagger}_{10}\|)$. Then contiguous sub-sequences of length $N_{min}$ elements will be chosen uniformly at random from all $N^{\dagger}_i$ for the final comparison. This is important, as longer sequences have more opportunities to build a comprehensive CFG. Table \ref{tab: surr_test_results} also depicts the value of $N_{min}$ for every \textit{r\=aga} pool.  If we choose $N_{min}$ to be constant over all the pools, then an insight about the overall complexity of compositions in the pool can be gathered by observing the time taken for compression ($\overline{t}_{calc}^{(n)}$, where $n$ is the number of surrogates.)

As a result, we end up introducing non-determinism to the equation, for we do not control which sub-sequence gets selected. Consequently, the results of LZP Causality are different over different iterations. In Table \ref{tab: surr_test_results}, the rows with number of surrogates $s = 0$  show the results of causal analysis amongst the compositions of the dataset (refer Table \ref{tab: raga_data}) using this methodology.

\section{Results}

In Tables \ref{tab: surr_test_results} and \ref{tab: surr_res_unif}, $\mathcal{E}_{s}$, where $s$ is the number of surrogates generated, stands for set of edges connecting a \textit{M\=e\d{l}akarta} composition and a \textit{Janya} composition. The results were calculated over \textbf{10} iterations, and as expected, they were not the same at every iteration. This is because:
\begin{enumerate}
    \item Different minimum-length sequences get selected at every iteration.
    \item Different surrogates are generated at every iteration.
\end{enumerate}

$N_{min}$ was defined previously in Section \ref{sec: surr_data_proc}, and it defines the length of all the sequences in the given \textit{r\=aga} pool.   

A correct connection is an edge that points from a \textit{M\=e\d{l}akarta} composition to a \textit{Janya} composition. Then $\mathcal{E}' \in \mathcal{E}$ is the subset of edges that point correctly. The mean cardinality of this set over 10 iterations is depicted as $\overline{\mathcal{E}'_{s}}$ in Table \ref{tab: surr_test_results}. 

The quantity $\overline{\mathcal{E}'_{s}}\%$ in Tables \ref{tab: surr_test_results} and \ref{tab: surr_res_unif} represents the \textbf{causal accuracy}: it is the percentage of the ratio of $\overline{\mathcal{E}'_{s}}$ and $\mathcal{E}_{s}$.

\begin{align}
    \label{eq: causal_accuracy}
    \overline{\mathcal{E}'_{s}}\% = \frac{\overline{\mathcal{E}'_{s}}}{\mathcal{E}_{s}} \times 100\%
\end{align}

The vector $\textbf{t}_{gen}^{(s)}$ contains the amount of time it takes to generate $s$ surrogates for each $\mathcal{R}_{\eta_1}$ and $\mathcal{R}_{\eta_2}^{\alpha}$ pair at every iteration. This value is expected to be stochastic in nature, primarily controlled by $\mathbbm{2}$ (Equation \eqref{eq: dur_dist}). The average over 10 iterations is depicted as $\overline{\textbf{t}}_{gen}^{(s)}$ in Table \ref{tab: surr_test_results}. The vector $\textbf{t}_{calc}^{(s)}$ contains the amount of time it takes to calculate the causality of the set $\{\mathcal{R}_{\eta} \cup \mathcal{R}_{\eta}^{\alpha}\}$ at every iteration. The average over 10 iterations is depicted as $\overline{\textbf{t}}_{calc}^{(s)}$. Note that all the experiments were performed on AMD's Ryzen 9 3950x, wherein, the surrogate generation was done on a single thread, and all the 32 threads were used for causal discovery.

\begin{table*}[!h]
    \caption{Results of Causal Analysis}
    \begin{tabular}{||>{\raggedright}p{1cm}|c|c|c|c|c|c||}
    \hline \hline
    \textbf{\textit{R\=aga} pool} & $\{\mathcal{R}_{8} \cup \mathcal{R}_{8}^{(d)}\}$ & $\{\mathcal{R}_{15} \cup \mathcal{R}_{15}^{(m)}\}$ & $\{\mathcal{R}_{22} \cup \mathcal{R}_{22}^{(a)}\}$ & $\{\mathcal{R}_{28} \cup \mathcal{R}_{28}^{(k)}\}$ & $\{\mathcal{R}_{29} \cup \mathcal{R}_{29}^{(h)}\}$ & $\{\mathcal{R}_{65} \cup \mathcal{R}_{29}^{(h)}\}$ \\ \hline
  $N_{min}$ & $84960$ & $43680$ & $112240$ & $98556$ & $84960$ & $114620$ \\ \hline
    $\mathcal{E}_{0}$ & $40$ & $24$ & $18$ & $20$ & $24$ & $24$ \\ \hline
   $\overline{\mathcal{E}'_{0}}$ & $10.4$ & $23.38$ & $16.72$ & $4.48$ & $21.58$ & $18.78$\\ \hline 
  $\overline{\mathcal{E}'_{0}}\%$ & $21.66$ & $97.04$ & $92.88$ & $22.4$ & $89.91$ & $78.25$\\ \hline
   $\mathcal{E}_{50}$ & $3248$ & $3024$ & $2968$ & $2970$ & $3024$ & $3024$ \\ \hline
   $\overline{\mathcal{E}'_{50}}$ & $340.1$ & $2389.1$ & $2083.8$ & $1214.3$ & $2630.6$ & $1927.2$ \\ \hline 
   $\overline{\mathcal{E}'_{50}}\%$ & $10.47$ & $79.0$ & $70.21$ & $40.89$ & $86.99$ & $63.73$  \\ \hline 
   $\mathcal{E}_{100}$ & $11448$ & $11024$ & $10918$ & $10920$ & $11024$ & $11024$ \\ \hline
   $\overline{\mathcal{E}'_{100}}$ & $976$ & $8522.7$ & $7798.3$ & $4559.4$ & $9934.1$ & $7687.$ \\ \hline 
   $\overline{\mathcal{E}'_{100}}\%$ & $8.53$ & $77.31$ & $71.43$ & $41.75$ & $90.11$ & $71.39$\\ 
    \hline \hline
    \end{tabular}
    \label{tab: surr_test_results}
\end{table*} 

As Table \ref{tab: surr_test_results} shows, $N_{min}$ plays an important role for obtaining the results. A larger value of $N_{min}$ presents LZ with a longer data stream, which results in a more accurate compression, however, it also directly affects $\overline{\textbf{t}}_{calc}^{(s)}$. The next logical step was to use a constant $N_{min} $ on all the \textit{r\=aga} pools. The results are depicted in Table \ref{tab: surr_res_unif}. Now, $\overline{\textbf{t}}_{calc}^{(s)}$'s are aligned in a different order. The \textit{r\=aga} pool  $\{\mathcal{R}_{8} \cup \mathcal{R}_{8}^{(d)}\}$ (\textit{Hanumat\=odi - Dhanyasi}) is the most ancient pool of the lot, and effectively contains some of the most complex compositional forms. It would be natural if the Lempel-Ziv algorithm spends most of the time in that pool. Consequently, $\overline{\textbf{t}}_{calc}^{(s)}$ is a great indicator of the overall complexity of the \textit{r\=aga} pool. Moreover, $\overline{\mathcal{E}}'_{s}$ in Table \ref{tab: surr_res_unif} shows the effect of a more inaccurate compression by LZ.

\begin{table*}[!h]
\centering
    \caption{Results of Causal Analysis, with $N_{min} = 43680$ for all the \textit{r\=aga} pools}
    \begin{tabular}{||>{\raggedright}p{1cm}|c|c|c|c|c|c||}
    \hline \hline
    \textbf{\textit{R\=aga} pool} & $\{\mathcal{R}_{8} \cup \mathcal{R}_{8}^{(d)}\}$ & $\{\mathcal{R}_{15} \cup \mathcal{R}_{15}^{(m)}\}$ & $\{\mathcal{R}_{22} \cup \mathcal{R}_{22}^{(a)}\}$ & $\{\mathcal{R}_{28} \cup \mathcal{R}_{28}^{(k)}\}$ & $\{\mathcal{R}_{29} \cup \mathcal{R}_{29}^{(h)}\}$ & $\{\mathcal{R}_{65} \cup \mathcal{R}_{29}^{(h)}\}$ \\ \hline
   $\overline{\mathcal{E}'_{50}}$ & $529.3$ & $2358.6$ & $1866$ & $1337.2$ & $2376$ & $1716.9$ \\ \hline 
   $\overline{\textbf{t}}_{gen}^{(50)}$ & $94.12$ & $51.17$ & $43.46$ & $61.03$ & $116.35$ & $56.66$ \\ \hline
   $\overline{\textbf{t}}_{calc}^{(50)}$ & $197.82$ & $151.8$ & $163.07$ & $183.0$ & $172.83$ & $160.8$ \\ \hline
   $\overline{\mathcal{E}'_{50}}\%$ & $16.3$ & $78$ & $62.87$ & $45.02$ & $78.57$ & $56.78$ \\ \hline 
   $\overline{\mathcal{E}'_{100}}$ & $1568.2$ & $8232.8$ & $6749.5$ & $4657.4$ & $8332.0$ & $5957.2$ \\ \hline 
   $\overline{\textbf{t}}_{gen}^{(100)}$  & $160.90$ & $127.86$ & $139.93$ & $182.30$ & $157.9$ & $115.58$ \\ \hline
   $\overline{\textbf{t}}_{calc}^{(100)}$ & $713.93$ & $557.18$ & $601.73$ & $671.83$ & $631.51$ & $589.89$\\ \hline
   $\overline{\mathcal{E}'_{100}}\%$ & $13.7$ & $74.68$ & $61.82$ & $41.83$ & $75.58$ & $54.0$ \\ 
    \hline \hline
    \end{tabular}
    \label{tab: surr_res_unif}
\end{table*} 

A homogeneous pattern can be observed in results depicted across Tables \ref{tab: surr_test_results} and \ref{tab: surr_res_unif}. Causal inference patterns observed from these results are tabulated in Table \ref{tab: inferences}. The overall inference does not add up for the $\{\mathcal{R}_{8} \cup \mathcal{R}_{8}^{(d)}\}$ (\textit{r\=aga Hanumat\=odi--Dhanyasi} pair) and $\{\mathcal{R}_{28} \cup \mathcal{R}_{28}^{(k)}\}$ (\textit{r\=aga Harik\=ambh\=oji--K\=ambh\=oji} pair). However, four out of the six chosen \textit{Janaka--Janya} pairs show desired results: The \textit{Janaka r\=aga} are a structural cause of the \textit{Janya r\=aga}.    

\begin{table*}[!h]
\centering
    \caption{Causal inference observed from the results depicted in Table \ref{tab: surr_test_results} and Table \ref{tab: surr_res_unif}. The arrow shows the causal direction, i.e. the $\mathcal{R}$ towards which the arrow points is the effect while the $\mathcal{R}$ from which it originates is the cause. }
    \begin{tabular}{||>{\raggedright}p{1cm}|c|c|c|c|c|c||}
    \hline 
    \textbf{\textit{R\=aga} pool} & $\{\mathcal{R}_{8} \cup \mathcal{R}_{8}^{(d)}\}$ & $\{\mathcal{R}_{15} \cup \mathcal{R}_{15}^{(m)}\}$ & $\{\mathcal{R}_{22} \cup \mathcal{R}_{22}^{(a)}\}$ & $\{\mathcal{R}_{28} \cup \mathcal{R}_{28}^{(k)}\}$ & $\{\mathcal{R}_{29} \cup \mathcal{R}_{29}^{(h)}\}$ & $\{\mathcal{R}_{65} \cup \mathcal{R}_{29}^{(h)}\}$ \\ \hline
    & $\{\mathcal{R}_{8} \leftarrow \mathcal{R}_{8}^{(d)}\}$ & $\{\mathcal{R}_{15} \rightarrow \mathcal{R}_{15}^{(m)}\}$ & $\{\mathcal{R}_{22} \rightarrow \mathcal{R}_{22}^{(a)}\}$ & $\{\mathcal{R}_{28} \leftarrow \mathcal{R}_{28}^{(k)}\}$ & $\{\mathcal{R}_{29} \rightarrow \mathcal{R}_{29}^{(h)}\}$ & $\{\mathcal{R}_{65} \rightarrow \mathcal{R}_{29}^{(h)}\}$ \\ \hline
    \hline 
    \end{tabular}
    \label{tab: inferences}
\end{table*}

\section{Discussion}

Mathematically, it is possible to obtain $100\%$ causal accuracy for any \textit{Janaka--Janya r\=aga} pair. However, that case would demotivate the existence of the \textit{Janya r\=aga}, especially as we are comparing different sections of the compositions and using textual inputs.  The \textit{Janya r\=aga} exist because they have a unique flavour/express a different mood, as compared to the corresponding \textit{M\=e\d{l}akarta}. In some cases, they might exhibit a unique combination of two different \textit{M\=e\d{l}akarta} (e.g. \textit{r\=aga Dhanyasi}, as explained later). 


In case of $\{\mathcal{R}_{28} \cup \mathcal{R}_{28}^{(k)}\}$ (\textit{r\=aga Harik\=ambh\=oji--K\=ambh\=oji} pair), \textit{r\=aga K\=ambh\=oji} is a \textit{Bh\=a\d{s}\=anga r\=aga}. The presence of \textit{K\=akal\=i N\=i\d{s}a\=ada} ($N_3$) along with the \textit{Kai\d{s}ik\=i N\=i\d{s}ada} ($N_2$)  in \textit{r\=aga K\=ambh\=oji} affects the context free grammar of the \textit{r\=aga}. The notations do not explicitly distinguish between the two, but as shown in Section \ref{sec: auto_parse}, they can be detected through the occurrence of certain phrases. Both the $Ni$'s are naturally represented by different note-indices ($N_2 \rightarrow 10$, $N_3 \rightarrow 11$) in the twelve tonal scale. 

When the $Ni$'s were not explicitly mined, the results for $\{\mathcal{R}_{28} \cup \mathcal{R}_{28}^{(k)}\}$ (\textit{r\=aga Harik\=ambh\=oji--K\=ambh\=oji} pair) were  $\overline{\mathcal{E}'_{0}}\% \approx 44\%$, $\overline{\mathcal{E}'_{50}}\% \approx 32\%$ and $\overline{\mathcal{E}'_{100}}\% \approx 36\%$. By parsing the $N_3$ explicitly, we are effectively adding a symbol to the grammar of $\mathcal{R}_{28}^{(k)}$ which does not exist in $\mathcal{R}_{28}$. In an ideal scenario, one could expect a further decrease in the causal accuracy ($\overline{\mathcal{E}'_{n}}\%$'s). However, we observe a slight increase overall (refer Table \ref{tab: surr_test_results}). This means that \textit{r\=aga K\=ambh\=oji} compositions can be shown to be derived from \textit{r\=aga Harik\=ambh\=oji}, if we sieve off the \textit{bh\=a\d{s}\=a\.nga} phrases.  

Another possible reason for lower causality on $\{\mathcal{R}_{28} \cup \mathcal{R}_{28}^{(k)}\}$ can be attributed to the fact that \textit{r\=aga K\=ambh\=oji} is an ancient \textit{r\=aga}, which has been in usage much before the \textit{Janaka, r\=aga Harik\=ambh\=oji}. As a result, \textit{r\=aga K\=ambh\=oji} has experienced more musicological experimentation than its \textit{M\=e\d{l}akarta}. 

But that is not the case with $\{\mathcal{R}_{8} \cup \mathcal{R}_8^{(d)}\}$ (the \textit{r\=aga Hanum\=at\=odi}--\textit{Dhany\=asi} pair). \textit{R\=aga T\=odi} (here known as \textit{r\=aga Hanum\=at\=odi}) is an ancient \textit{r\=aga}, and has been extensively experimented upon by different composers of all the eras. However, observe the \textit{\=ar\=oha} sequence of $\mathcal{R}_{8}^{(d)}$ (\textit{r\=aga Dhanyasi}) in Table \ref{tab: raga_data}. It might as well be a derived from $\mathcal{R}_{22}$ (\textit{r\=aga Kharaharapriya}). In fact, \textit{r\=aga Dhanyasi} uses the \textit{\=ar\=oha} sequence of \textit{r\=aga \'Suddha Dhanyasi}, a \textit{Janya} of \textit{r\=aga Kharaharapriya} and the \textit{avar\=oha} sequence of \textit{r\=aga Hanumat\=odi}. Hence, LZP might struggle in $\{\mathcal{R}_{8} \cup \mathcal{R}_{8}^{(d)}\}$ because of the fact that \textit{r\=aga Dhanyasi} is a complex combination of two differently coloured \textit{r\=aga}. Consequently, LZP sees \textit{r\=aga Hanumat\=odi} as a ``purer/simpler'' scale and often ends up predicting it as an effect rather than a cause. Also, \textit{r\=aga T\=odi} and \textit{r\=aga Dhanyasi} are complex \textit{r\=aga}, and their renditions usually require heavy usage of embellishments and ornamentation, which are not accurately annotated.  

Moreover, not only $\mathcal{R}_{8}$ (\textit{r\=aga Hanumat\=odi}), but almost all the \textit{r\=aga} above suffer from the representational limitations. Representational limiters are explained in detail below. 


\subsection{Representational Limitations}
\label{sec: rep_lim}
Representational limitations refer to the shortcomings of using text as a representation. A primary example of representational limitations are the \textit{gamaka}. \textit{Gamaka} are complex ornamented glides that revolve around the central frequency of a \textit{s\'vara}. There are \textit{da\d{s}avidha gamaka} i. e. ten different styles of \textit{gamaka}. There are massive differences in rendering style of these \textit{gamaka} amongst different traditions. Naturally, they are very difficult to annotate. As such, text representations of Carnatic compositions seldom contain any \textit{gamaka} references. These musical embellishments usually are taught orally to the students of Carnatic music. In many \textit{r\=aga} of Carnatic music, \textit{gamaka} are inherent to the existence of the \textit{r\=aga}. For example, \textit{r\=aga Sah\=an\=a} cannot exist without its \textit{gamaka}.

One of the reasons why \textit{r\=aga Hanumat\=odi--Dhanyasi} $\{\mathcal{R}_{8} \cup \mathcal{R}_{8}^{(d)}\}$ might fail is because of the fact that text representation of compositions does not capture \textit{gamaka}. \textit{R\=aga Hanumat\=odi}, which is more popular by the name \textit{r\=aga T\=odi}, gives importance to every note. In typical \textit{r\=aga T\=odi} renditions, every note has its own peculiarity and unique usage in each phrase. Representing these peculiarities in text is a tough task.
 
Another example of a representational limitation is that while using text notations, distinguishing between the \textit{\=ar\=oha} and \textit{avar\=oha} phrases becomes difficult. Quite often, text representations are not presented in the \texttt{UTF-8} rich text format, as a result of which, $\dot{S}$ gets represented just as $S$, and as a result, octave resolution is annulled. This problem and it's workaround have been discussed previously in Section \ref{sec: auto_parse}. However, this representational limitation only reinforces the difficulty of \textit{\=ar\=oha - avar\=oha} distinction. \textit{\=Ar\=oha--avar\=oha} phrase distinction means the ability to evaluate the direction of the current phrase, and the ability to predict the direction upcoming phrases. 



\subsection{Parse Limitations}
\label{sec: parse_lim}

Parser limitations refer to the shortcomings of the parser described in Section \ref{sec: auto_parse} and Appendix \ref{app: auto-parser}. 

In Carnatic music, a composition is divided into several sections. 
The reason behind this division is that, the \textit{r\=aga} needs to be exposed gradually, while rendering, for the best effect. However, the parser does not guarantee this structural bifurcation. Moreover, as explained in Section \ref{sec: data_proc}, a random section of the composition is considered for the final causal inference. This, in the worst case, means that a fast moving section (e.g. \textit{cara\d{n}a}) of one composition might get contested against a more serene section (e.g. \textit{pallavi}) of another. Hence, even if the former is a \textit{Janya} composition, \texttt{LZP} will tend to evaluate it as the cause.

Naturally, the length of each section varies for every composition. The primary reason why the bifurcation of these sections is lost is because the parse output produces a sequence of note events as shown in Equation \eqref{eq: n_dagger}. Moreover, depending upon the chosen $N_{min}$, a variable number of sections might get selected. This might lead to further complications. For example, consider a scenario where the \textit{pallavi, anupallavi} are selected along with a very small portion of the \textit{cara\d{n}a}. Then this very small \textit{cara\d{n}a} portion might add unnecessary complexity for \texttt{LZP}.


As highlighted in Section \ref{sec: auto_parse}, as long as notations are not represented in a \texttt{utf-8} format, absolute resolution of the length (in time) of each note-event (as represented $\hat{b}$ in Equation \eqref{eq: 3d-space}) and the octave of each \textit{s\'vara} are non-trivial problems. Because of the way the parser works, we quite often encounter situations where we reach notes like $\underdot{S}$ or $\dot{N}$ (outside the $[\underdot{P}, \dot{P}]$ interval), which are seldom used in any Carnatic music rendition. Such notes add faux vocabulary, which increase the \texttt{LZ} complexity of the compositions, making them more complex than they actually are. This also affects the stationary distribution ($\pi$) of the \textit{r\=aga}. Consequently, during surrogate generation, sometimes a problem is encountered when the randomly selected pitch belongs to this faux vocabulary. These notes act as "traps" for surrogate generation because they might transition only to themselves or to one other pitch, which in turn transitions back to the original faux pitch leading to a vicious cycle\footnote{An example for the latter case: consider a scenario where the parser has erratically decoded the last phrase of a particular composition as $P \dot{D} \dot{N} \dot{D}$, and that those are the only occurrences of $\dot{N}$ and $\dot{D}$ in the whole \textit{r\=aga}. Now, imagine that, while surrogate generation, a $\dot{D}$ gets selected. That $\dot{D}$ can only transition to the $\dot{N}$, which in turn only transitions to the $\dot{D}$. That leads to a trap. }. To tackle this challenge in surrogate generation, whenever a faux pitch is encountered, the parser either tries to move towards the real vocabulary, and if that's not possible, the whole note-event is discarded and the generation starts over.

In typical Carnatic notations (refer Figure \ref{fig: representation}), when a group of \textit{s\'vara} are clubbed together, a Carnatic practitioner usually understands that all those notes need to be sung together to fit a specific number of counts. Unfortunately, the specific number of counts always varies. That is why for a given parsed note event $\vec{n}$ (refer Equation \eqref{eq: 3d-space}), many times $n_b$ is not accurate. However, the workaround discussed in Section \ref{sec: auto_parse} works as a good approximation. 



\subsection{Future Directions}

One way to attack the problems highlighted in Sections \ref{sec: rep_lim} and \ref{sec: parse_lim} is to use an audio of several renditions along with the text notations. Analysing the contours in a Fourier Transform of the audio will help in detecting \textit{gamaka}. These audio files can then be used to train specialized architectures like the auto-encoder network used in \cite{dhariwal2020jukebox} to find a more elegant representation for each \textit{r\=aga}. Historically, some ancient \textit{r\=aga} have existed long before their classification as a \textit{Janya} of some specific \textit{M\=e\d{l}akarta} in the $17^{th}$ C.E. Specialized architectures can be trained to learn different features of each \textit{r\=aga}, and these features can hopefully be used for counterfactual reasoning of the historical causal events. 
           
Augmenting the current text dataset with audio renditions can also help us redefine the concept of \textit{r\=aga} in a more abstract manner. This abstract definition might help in reducing ambiguities inherent to Carnatic music, especially ones which cause difficulties in its analysis.  For example, in Carnatic practice, the rendition protocol of \textit{r\=aga Ha\.msadhwani} is often considered to be similar to that of \textit{r\=aga Kaly\=a\d{n}i}, even though it is considered a \textit{Janya} of \textit{r\=aga \'Sa\.nkar\=abhara\d{n}a\.m}. As we have shown in this article, causal analysis might be able to answer that conundrum. Redefinition of the concept of \textit{r\=aga} using hybrid data might aid in solidifying our claim. 

The use of hybrid data can also help capture nuances in each note of a \textit{r\=aga T\=odi} rendition. Because of this new ability, the true complexity of \textit{r\=aga T\=odi} can be highlighted, which may improve the accuracy of causal discovery. 
           
Carnatic music is constantly evolving, and hence, a full causal analysis of the subject is next to impossible. But we believe that causal analysis can act as a useful tool for studying the past and present evolution of the subject. In this regard, our study has initiated this line of causal analysis of Carnatic music for the very first time.

\section{Conclusions}

Carnatic music is a huge mine of data, but a large part of that tends to be inaccessible for analysis. This is primarily because modern Carnatic music stands firmly on the works of the Trinity\footnote{The Trinity refers to the three most prolific Carnatic composers of the $18^{th}$ century, \'Syama \'Sastri, Ty\=agar\=aja Sw\=ami and Muttusw\=ami D\=ik\d{s}itar.}, and other well-known composers of the the same and previous era. As a result, even though the scientific principles on which it is based are solid, the number of compositions in practice are limited. This is one of the main disadvantages of the oral tradition. 

Knowledge in traditional disciplines has been orally passed across generations in our country, usually in the form of \textit{\'sruti} (here, listening) and \textit{smriti} (here, remembering). And so has it been in the case of music, which has had a rich and traditional \textit{gur\=u--\'si\d{s}ya parampar\=a}, wherein the \textit{gur\=u} or teacher imparts knowledge or information to the \textit{\'si\d{s}ya} or disciple. These sources of knowledge are orally learnt by the disciple, memorised and passed on to the next generation. In case of written texts, most of our knowledge was written on perishable sources such as palm leaves and paper manuscripts. Most of these sources did not survive due to the fragile nature of the materials used, leading to the loss of documented knowledge. As a result, no specific rule-book exists for Carnatic music. Every school tends to follow their own practices, which might vary accordingly. 

This article gives an insight into how the \textit{M\=e\d{l}akarta r\=aga} are structurally related to their \textit{Janya r\=aga}. These causal relationships are a step towards re-discovering the lost scientific principles on which the whole system of Carnatic music is founded. From the evolution of concepts of \textit{gr\=ama} and \textit{m\=urchana} to the concept of the \textit{r\=aga}, which itself went through innate changes in every century, a clear causal path can be traced. Hence, on further development, causal analysis of Hindustani and Carnatic music together can be employed for finding the insights behind the genesis of the \textit{r\=aga} itself.

\appendix
\section{The auto-parser protocol}
\label{app: auto-parser}
The auto-parser featured in this article, primarily in Section \ref{sec: auto_parse} is a simple automatic text-to-text converter, which takes the \textit{s\'vara} section of the text (composition) as the input and output their representation in the vector space $\mathbf{S}$ (Equation \eqref{eq: 3d-space}). The following steps are followed:
\begin{itemize}
    \item The \textit{s\'vara} are stored in a plain-text file, which is visible as a stream of bytes to the parser. The parser reads this stream line by line, where a line is characterized by the newline character (typically \texttt{\textbackslash n}). Each line is then analysed from left to right. 
    \item The parser essentially needs to output $\textbf{N} := \{\Vec{n_1}, \Vec{n_2}, \ldots, \Vec{n_K}\}, \Vec{n_i} \in \mathbf{S}$ (Equation \eqref{eq: 3d-space}), and $i \in [1:K]$, which is the set of all the note-events that are annotated in the text. Notice that the sequence $\textbf{N}_c = \{(n_1)_c, (n_1)_c, \ldots, (n_K)_c\}$ (the sequence of all the coefficients of $\hat{c}$ throughout the composition) is a monotonically non-decreasing sequence, which gets incremented by $1$ every time the double \textit{da\d{n}\d{d}\=a} ($||$) is seen. Hence, by default, while the first character of the first line is being analysed, the $\hat{c}$ coefficient of that event is set to zero ($(n_1)_c = 0$). 
    \item One by one, each character in the line is analysed. As explained in Section \ref{sec: auto_parse}, $(n_i)_a, ~\forall ~i\in[1:K]$ is set based on the observed \textit{s\'vara}. The parser has access to a database that contains the vocabulary of each \textit{r\=aga} listed in Table \ref{tab: raga_data}. By default, if the \textit{r\=aga} of the current composition is found in the database, each $(n_i)_a$ is set according to the twelve-tonal scale, otherwise the seven-tonal scale is used. Sometimes $(n_i)_a$ may also get affected by the neighbouring events. Section \ref{sec: kambhoji} highlights one such example, where $(n_{i-1})_a$ gets affected by $(n_{i-2})_a$ and $(n_{i})_a$, for very specific values of both $(n_{i-2})_a$ and $(n_{i})_a$. 
    \item Similarly, $(n_i)_b, ~\forall ~i\in[1:K]$ is set based on the letter case of the detected character, and may get affected at the occurrence of the double \textit{da\d{n}\d{d}\=a} (again, as explained in Section \ref{sec: auto_parse}).
    \item The final $\textbf{N}$ is then output as comma-separated values.
\end{itemize}

In this manner, the parser outputs the three-dimensional representation of the composition, which is ready for further processing explained in Sections \ref{sec: data_proc} and \ref{sec: surr_test}.

\section{The Lempel-Ziv Penalty}

The \texttt{Lempel-Ziv Penalty} framework for causal inference has been explained in detail in \cite{pranay2021causal}. This section provides a brief introduction to the framework. 

Context-free grammar (CFG) of a sequence is the grammar inferred by a lossless compression scheme for compressing the data. The CFG may be unique for every sequence. Consider a compression-complexity measure $\mathcal{C}$. Define $G$, the CFG of a sequence $x$ such that the following optimization is satisfied:
\begin{align}
    \label{eq: cfg}
    \min \|\mathcal{C}(x|G_x)\|,
\end{align}
which implies that the CFG of x $G_x$ is a set which minimizes the cardinality of $\mathcal{C}(x)$. 

Consider two sequences $x$ and $y$. The \textit{penalty} framework helps us evaluate the causal relations between $x$ and $y$. Let $G_x$ be the CFG of $x$ and $G_y$ be the CFG of $y$.  Then, the penalty incurred from using $G_x$ to compute $\mathcal{C}(y)$ is given as
\begin{align}
    \label{eq: penalty_1}
    P(x \rightarrow y) &= \mathcal{C}(y|G_x) - \mathcal{C}(y),
\end{align}
and vice-versa ($P(y \rightarrow x) = \mathcal{C}(x|G_y) - \mathcal{C}(x)$). The \textit{penalty} framework then suggests that if $P(x \rightarrow y) < P(y \rightarrow x)$, then $x$ is the causal sequence and $y$ is the effect, and vice-versa. In case $P(x \rightarrow y) = P(y \rightarrow x)$, then the sequences $x$ and $y$ are either causally unrelated, or mutually independent. 

In the \texttt{Lempel-Ziv Penalty} framework, $\mathcal{C}$ is the Lempel-Ziv Complexity (LZ76) \cite{lempel1976complexity}. Because LZ76 views the sequence to be compressed as a stream of data from left to right, if we concatenate $y$ to the right of $x$, then $\mathcal{C}(xy) = \mathcal{C}(y|G_x)$.

\section*{Acknowledgment}

The authors thank Indian Heritage in Digital Space (IHDS), Interdisciplinary Cyber Physical Systems (ICPS) Programme of the Department of Science and Technology, Government of India for funding this study, which forms part of the project ``Early Fusion Music: Cross-Cultural Musical Exchanges in Colonial India from the Late $18^{th}$ to the Early $20^{th}$ Century'' (Sanction No DST/ICPS/IHDS/2018 (General), dated 13 March 2019).

The authors also thank Pranay S Yadav for the Python implementation (\texttt{ETCPy}) of the \texttt{Lempel-Ziv Penalty} model for causal discovery.

\bibliographystyle{unsrt}
\bibliography{main}

\begin{landscape}
\section*{Supplementary Material}
\subsection*{Causal Inference for all \textit{M\=e\d{l}akarta--Janya r\=aga} pairs} 

This section depicts typical results of Causal inference in all the \textit{M\=e\d{l}akarta--Janya r\=aga} pairs.  As described in the article, the results of causal discovery in each \textit{M\=e\d{l}akarta--Janya} pair can be visualized in the form of a Directed Acyclic Graph. Outgoing arrows from a node make it a cause, and likewise, incoming arrows make it an effect.  

\begin{itemize}
    \item Figure \ref{fig: caus_graph_8} shows a causal inference DAG for the \textit{r\=aga Hanumat\=odi--Dhanyasi} pair.
    \item Figure \ref{fig: caus_graph_15} shows a causal inference DAG for the \textit{r\=aga M\=ayama\b{l}avagau\b{l}a--Malahari} pair.
    \item Figure \ref{fig: caus_graph_22} shows a causal inference DAG for the \textit{r\=aga Kharaharapriya--\=Abh\=ogi} pair. This DAG has been structured in the Left-Right direction, i.e. leftmost nodes are the ultimate causes and the rightmost nodes are the ultimate effects.
    \item Figure \ref{fig: caus_graph_28} shows a causal inference DAG for the \textit{r\=aga Harik\=ambh\=oji--K\=ambh\=oji} pair.
    \item Figure \ref{fig: caus_graph_29} shows a causal inference DAG for the \textit{r\=aga \'Sankar\=abhara\d{n}a\.m--Ha\d{m}sadhwani} pair.
\end{itemize}
\begin{figure*}[!hbt]
    \includegraphics[scale=0.112]{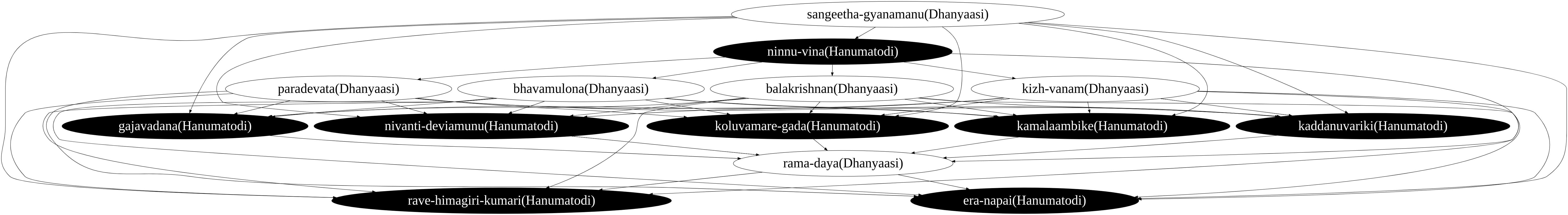}
    \caption{Result of LZP causal analysis on $\{\mathcal{R}_{8} \cup \mathcal{R}_{8}^{(d)}\}$. Black background highlights the \textit{M\=e\d{l}akarta} compositions.}
    \label{fig: caus_graph_8}
\end{figure*}
\begin{figure*}[!hbt]
    \centering
    \includegraphics[scale=0.2]{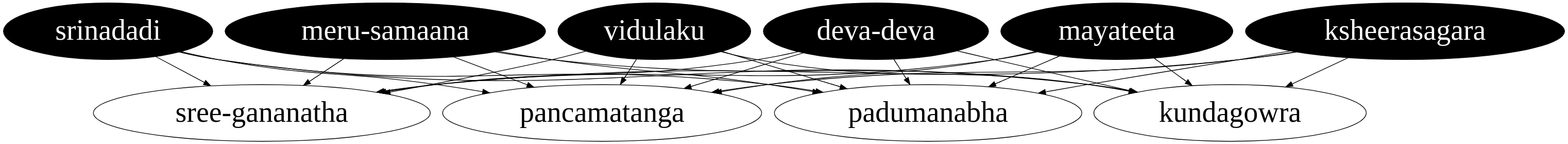}
    \caption{Result of LZP causal analysis on $\{\mathcal{R}_{15} \cup \mathcal{R}_{15}^{(m)}\}$. Black background highlights the \textit{M\=e\d{l}akarta} compositions.}
    \label{fig: caus_graph_15}
\end{figure*}

\end{landscape}

\begin{landscape}
\begin{figure*}[!hbt]
    \centering
    \includegraphics[scale=0.15]{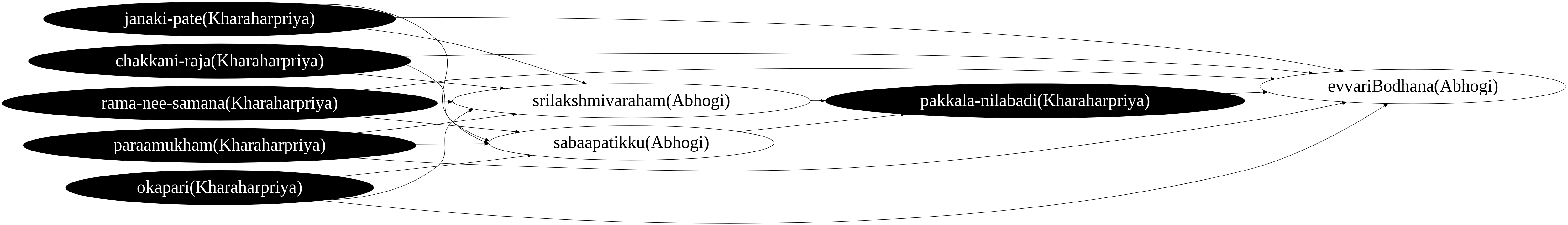}
    \caption{Result of LZP causal analysis on $\{\mathcal{R}_{22} \cup \mathcal{R}_{22}^{(a)}\}$. Black background highlights the \textit{M\=e\d{l}akarta} compositions. This DAG has a Left-Right structure instead of a Top-Down structure.}
    \label{fig: caus_graph_22}
\end{figure*}
\begin{figure*}[!hbt]
    \centering
    \includegraphics[scale=0.2]{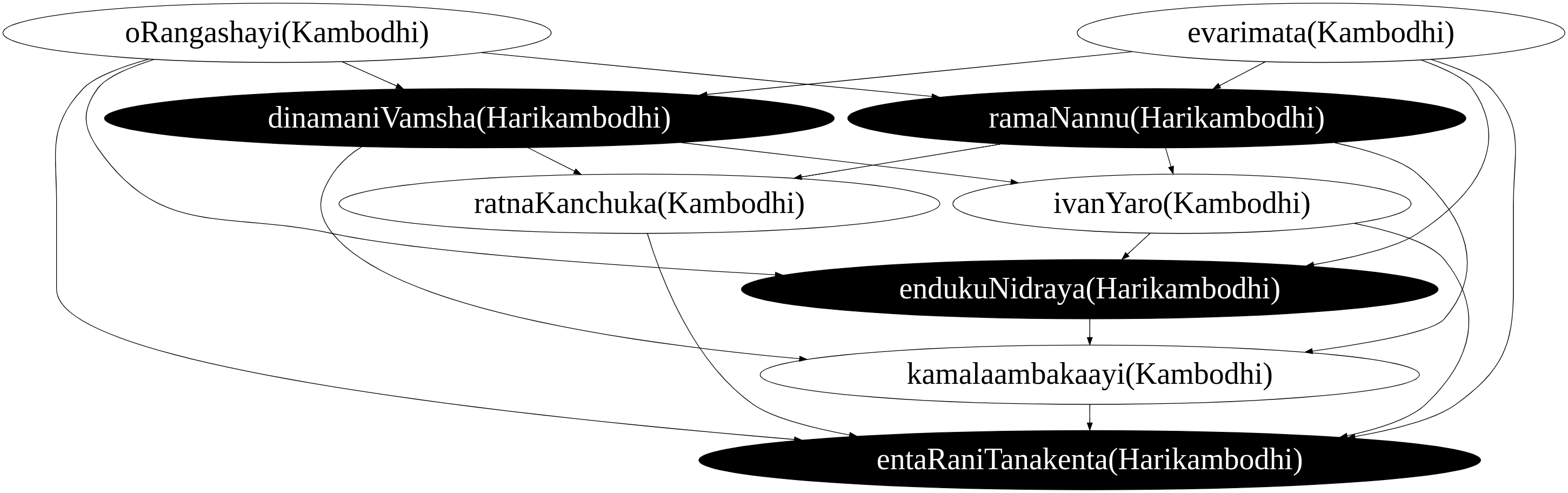}
    \caption{Result of LZP causal analysis on $\{\mathcal{R}_{28} \cup \mathcal{R}_{28}^{(k)}\}$. Black background highlights the \textit{M\=e\d{l}akarta} compositions.}
    \label{fig: caus_graph_28}
\end{figure*}
\end{landscape}

\begin{landscape}
\begin{figure*}[!hbt]
    \centering
    \includegraphics[scale=0.175]{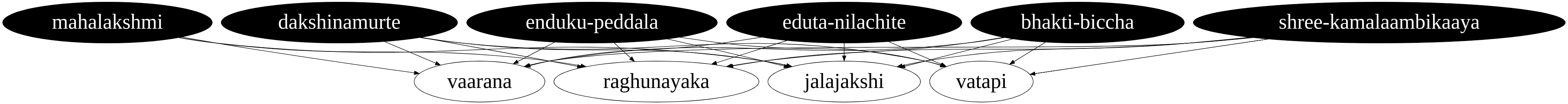}
    \caption{Result of LZP causal analysis on $\{\mathcal{R}_{29} \cup \mathcal{R}_{29}^{(h)}\}$. Black background highlights the \textit{M\=e\d{l}akarta} compositions.}
    \label{fig: caus_graph_29}
\end{figure*}
\subsection*{Stationary Distribution of each \textit{r\=aga}}
In this section, the stationary distribution ($\pi(\mathcal{R})$) of each \textit{r\=aga} along with that of its surrogate compositions ($\pi(\Phi(\mathcal{R}))$) is depicted, $\forall ~\mathcal{R} ~\in \mathbf{C}$, where $\mathbf{C}$ is the set of all \textit{r\=aga} in the dataset. The Y-axis in these plots represents probability of the occurrence of the note corresponding to the index depicted in the X-axis. Notice that the zero of the X-axis does not correspond to a pitch index of zero, instead corresponds to the lowest pitch index found in that \textit{r\=aga}. The last point in each plot gives the probability of occurrence of rests (which were given a pitch-index of $\infty$).  
\begin{figure*}[!hbt]
\begin{subfigure}[b]{0.48\textwidth}
    \centering
    \includegraphics[width=0.9\columnwidth]{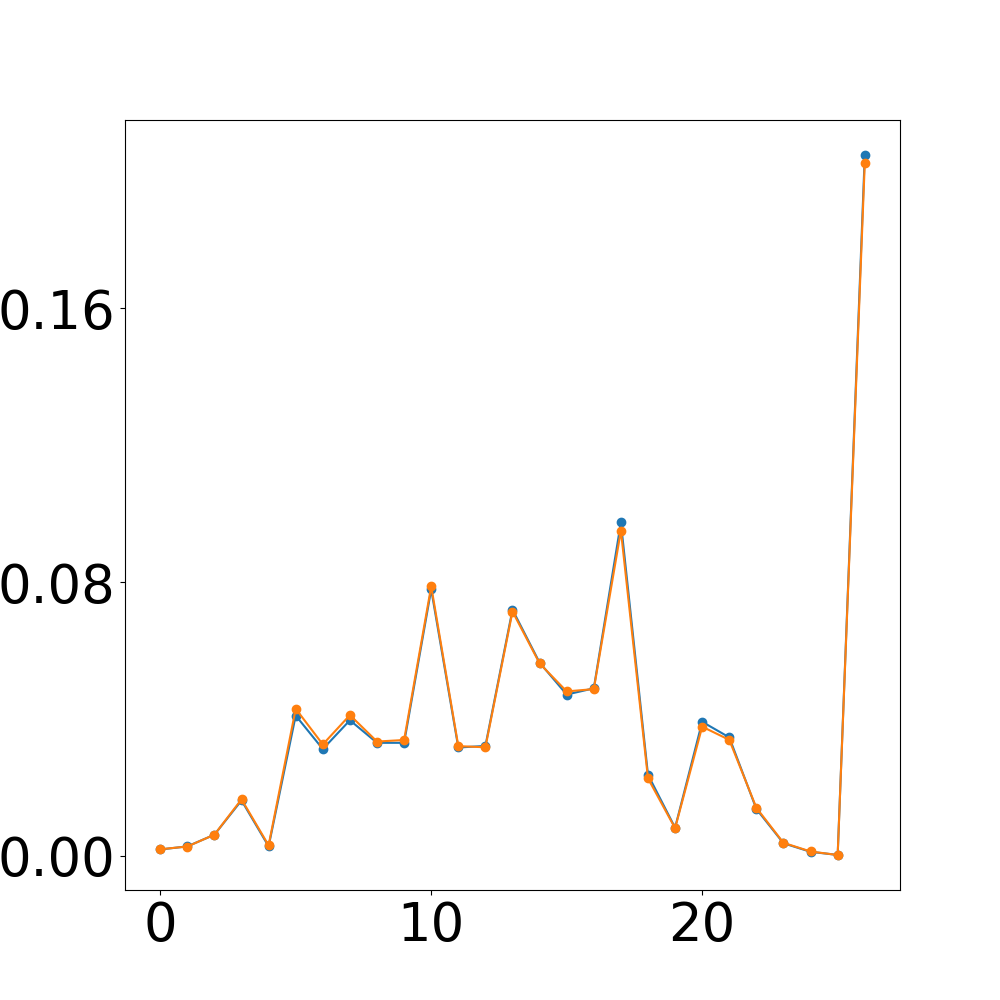}
    \caption{$\pi(\mathcal{R}_{8}^{(d)})$ (blue) and $\pi(\Phi(\mathcal{R}_{8}^{(d)}))$ (orange). }
    \label{fig: surr_stat_8_d}
\end{subfigure}
\begin{subfigure}[b]{0.48\textwidth}
    \centering
    \includegraphics[width=0.9\columnwidth]{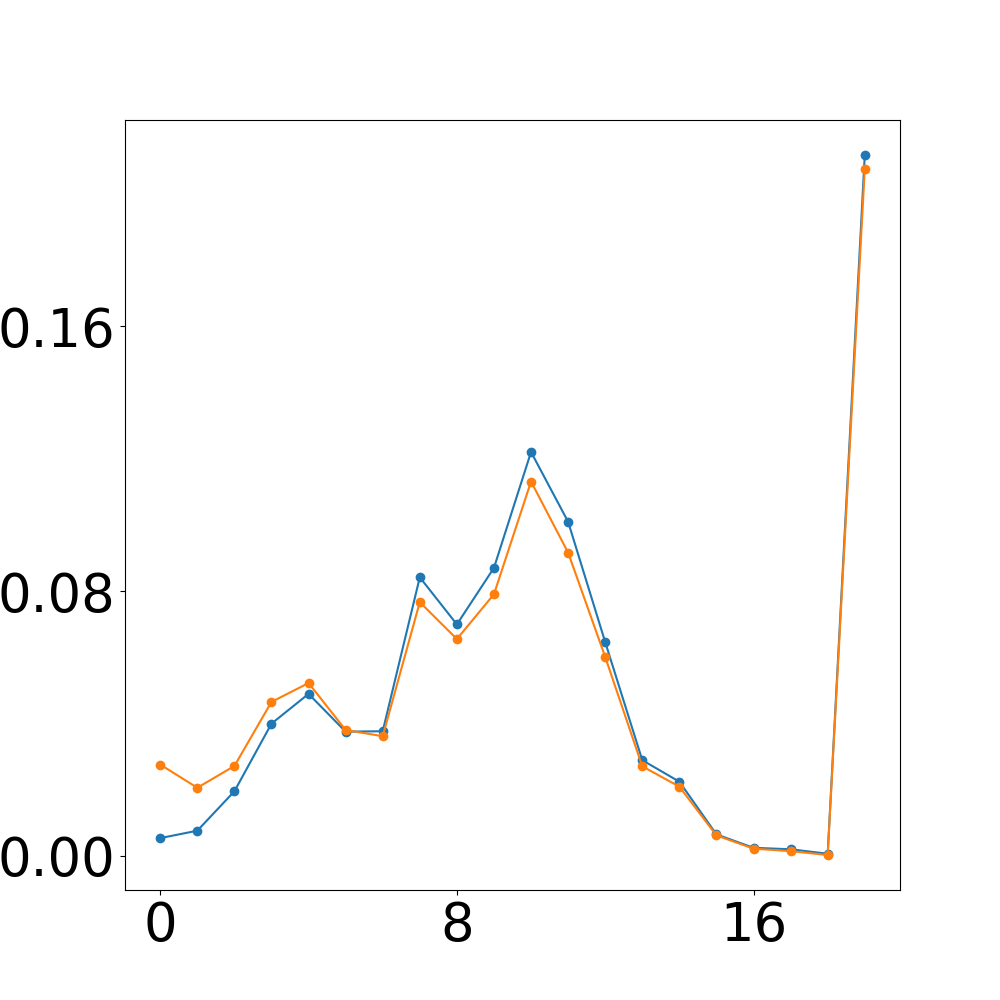}
    \caption{$\pi(\mathcal{R}_{15})$ (blue) and $\pi(\Phi(\mathcal{R}_{15}))$ (orange). }
    \label{fig: surr_stat_15}
\end{subfigure}
\begin{subfigure}[b]{0.48\textwidth}
    \centering
    \includegraphics[width=0.9\columnwidth]{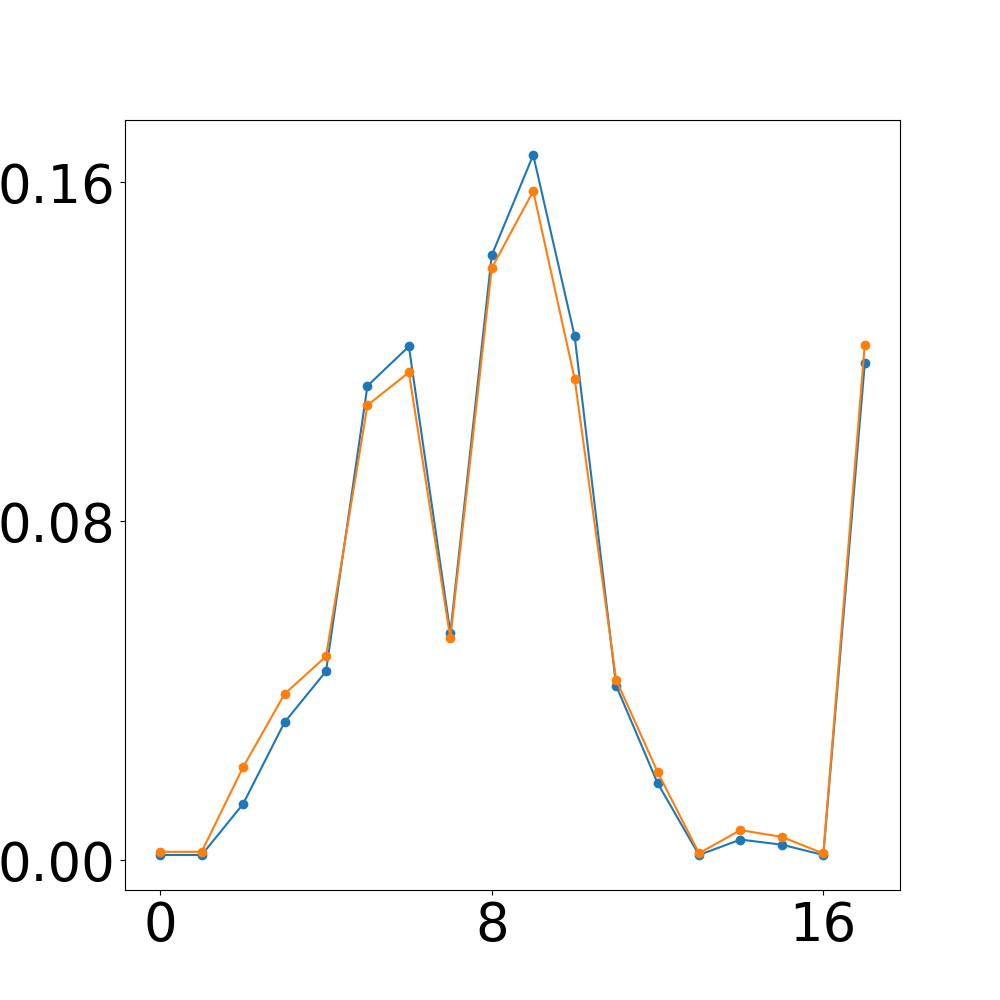}
    \caption{$\pi(\mathcal{R}_{15}^{(m)})$ (blue) and $\pi(\Phi(\mathcal{R}_{15}^{(m)}))$ (orange). }
    \label{fig: surr_stat_15_m}
\end{subfigure}
\caption{The stationary distribution of the specified \textit{r\=aga} compared to the stationary distribution of its surrogates.}
\label{fig: surr_statdists}
\end{figure*}

\end{landscape}
\begin{landscape}
In Figure \ref{fig: surr_stat_15_m}, rests are not the most frequently occurring element. This can be justified by having a look at the dataset of \textit{r\=aga Malahari}, which mostly contains short and succinct \textit{g\=itha}. These \textit{g\=itha} are usually used as beginner lessons, because a note event quite often lasts for one count, and the use of \textit{gamaka} is minimal.
\begin{figure*}[!hbt]
\begin{subfigure}[b]{0.48\textwidth}
    \centering
    \includegraphics[width=0.8\columnwidth]{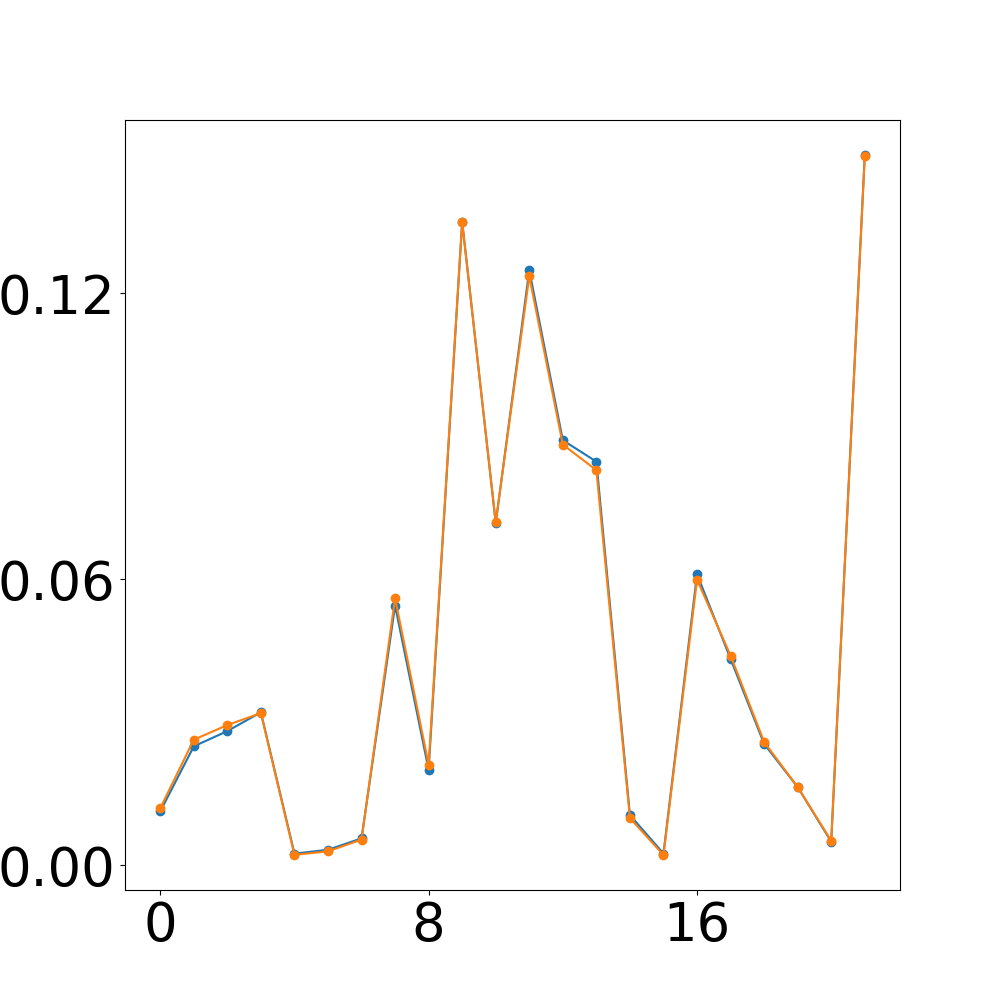}
    \caption{$\pi(\mathcal{R}_{22}^{(a)})$ (blue) and $\pi(\Phi(\mathcal{R}_{22}^{(a)}))$ (orange). }
    \label{fig: surr_stat_22_a}
\end{subfigure}
\begin{subfigure}[b]{0.48\textwidth}
    \centering
    \includegraphics[width=0.8\columnwidth]{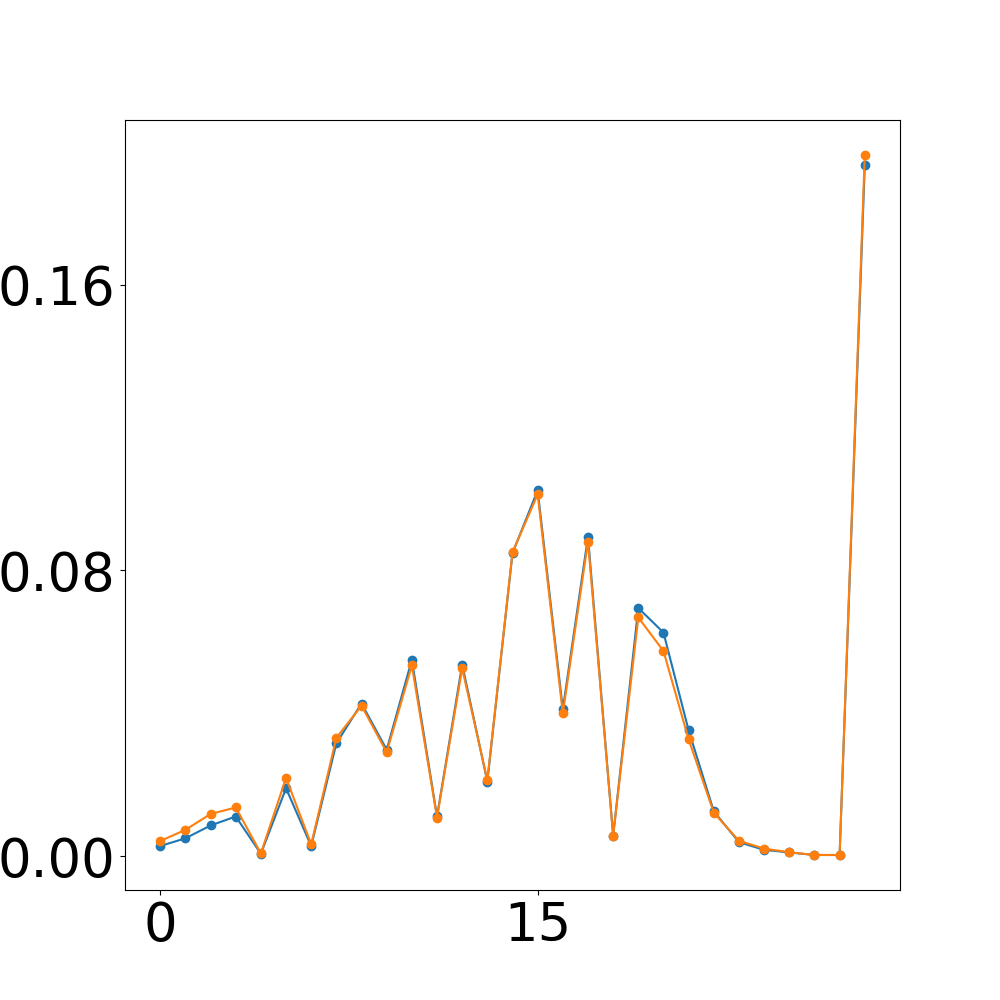}
    \caption{$\pi(\mathcal{R}_{28})$ (blue) and $\pi(\Phi(\mathcal{R}_{28}))$ (orange). }
    \label{fig: surr_stat_28}
\end{subfigure}
\begin{subfigure}[b]{0.48\textwidth}
    \centering
    \includegraphics[width=0.8\columnwidth]{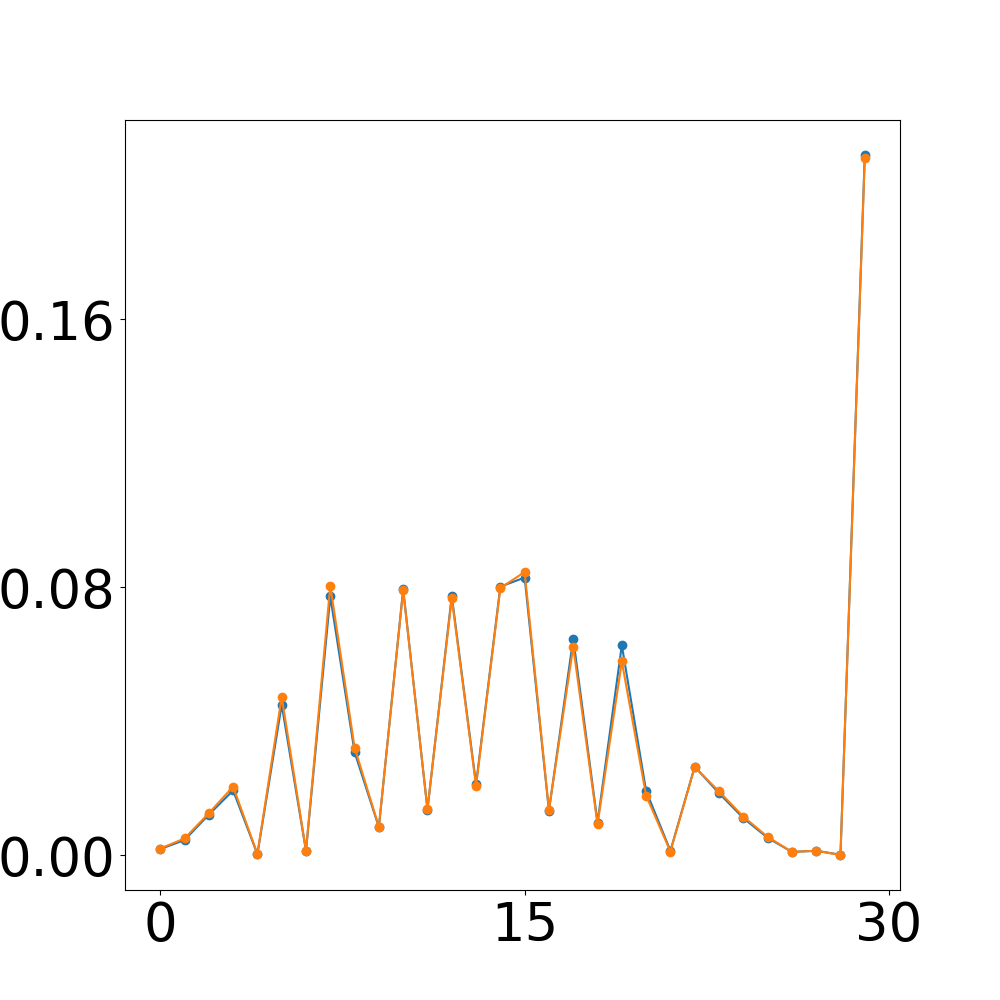}
    \caption{$\pi(\mathcal{R}_{28}^{(k)})$ (blue) and $\pi(\Phi(\mathcal{R}_{28}^{(k)}))$ (orange). }
    \label{fig: surr_stat_28_k}
\end{subfigure}
\begin{subfigure}[b]{0.48\textwidth}
    \centering
    \includegraphics[width=0.8\columnwidth]{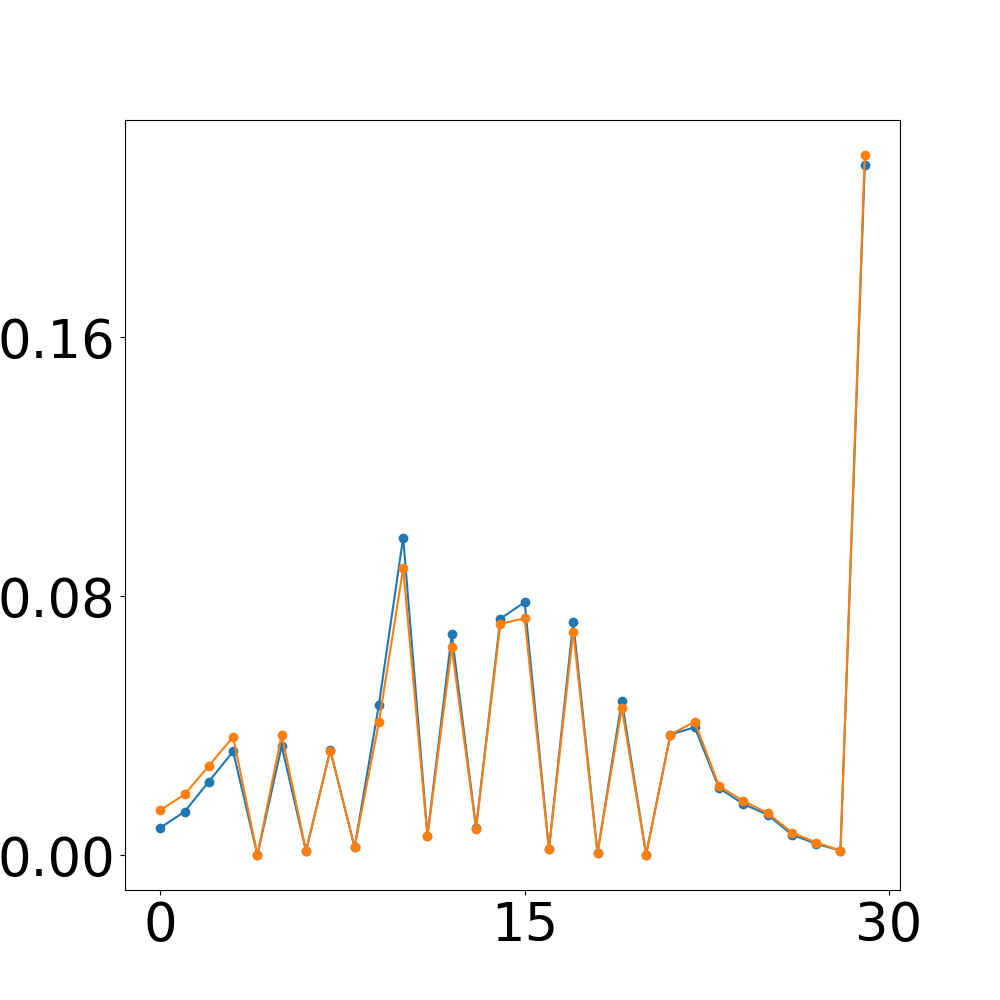}
    \caption{$\pi(\mathcal{R}_{29})$ (blue) and $\pi(\Phi(\mathcal{R}_{29}))$ (orange). }
    \label{fig: surr_stat_29}
\end{subfigure}
\begin{subfigure}[b]{0.48\textwidth}
    \centering
    \includegraphics[width=0.8\columnwidth]{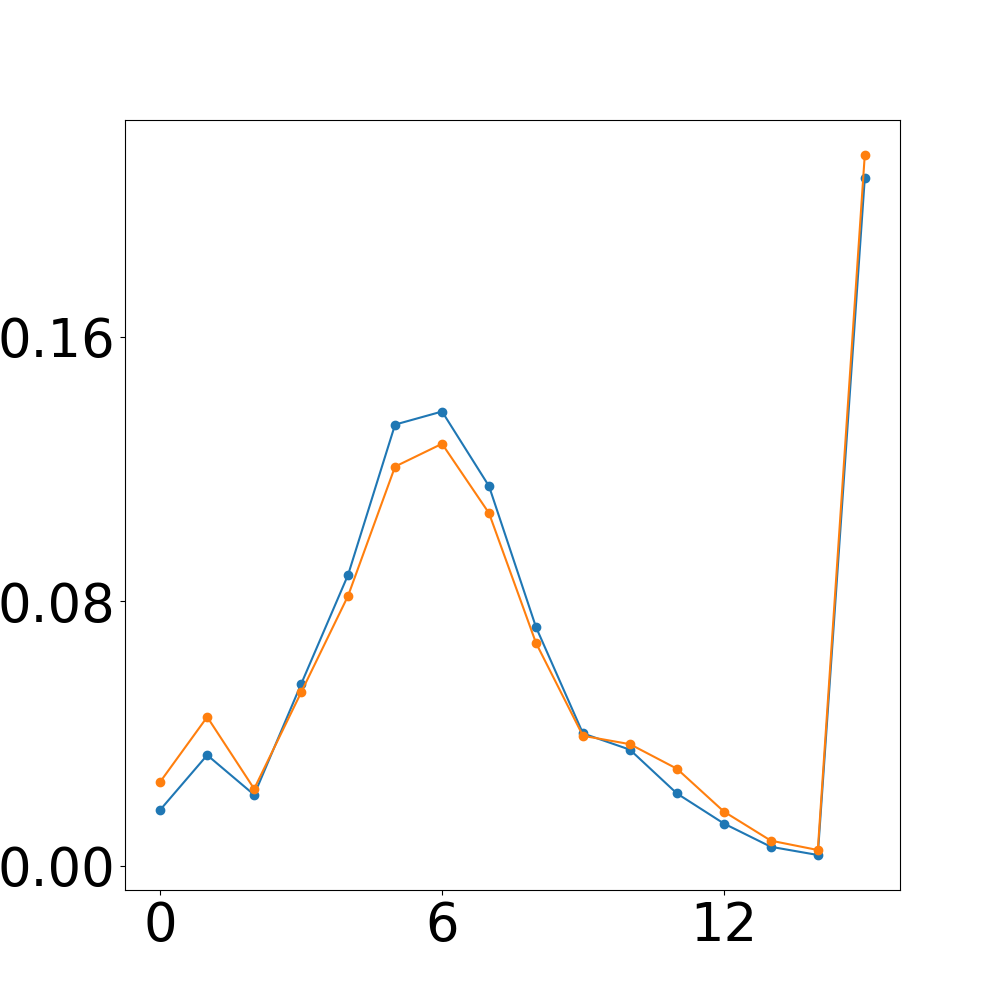}
    \caption{$\pi(\mathcal{R}_{29}^{(h)})$ (blue) and $\pi(\Phi(\mathcal{R}_{29}^{(h)}))$ (orange). }
    \label{fig: surr_stat_29_h}
\end{subfigure}
\caption{The stationary distribution of the specified \textit{r\=aga} compared to the stationary distribution of its surrogates.}
\label{fig: surr_statdists}
\end{figure*}
As discussed earlier, \textit{R\=aga K\=ambh\=oji} employs the \textit{K\=akali Ni\d{s}\=ada} ($N_3$), which explains Figure \ref{fig: surr_stat_28_k}'s similarity with Figure \ref{fig: surr_stat_29}. 
\end{landscape}

\end{document}